\documentclass[12pt]{article}

\usepackage{graphicx}
\usepackage{dcolumn}
\usepackage{bm}
\usepackage[            
pdftex,                 
pdfborder= 0 0 0,
a4paper,                
colorlinks=true,        
citecolor=green,        
linkcolor=cyan,         
menucolor=blue,         
urlcolor=magenta,       
bookmarksopen=true
]{hyperref}

\usepackage{amsmath}
\usepackage{amssymb}

\newcommand{\PNAS}{Proceedings of the National Academy of Sciences
                   of the United States of America}

\newcommand{\z}{Z\kern-0.45emZ}
\newcommand{\vi}{I\kern-0.3emB}
\newcommand{\1}{I\kern-0.3emI}

\newcommand{\e}{I\kern-0.3emE}
\newcommand{\E}{I\kern-0.3emE}
\newcommand{\ind}{I\kern-0.3emI}
\newcommand{\p}{I\kern-0.3emP}

\newcommand{\be}{\begin{equation}}
\newcommand{\ee}{\end{equation}}

\begin{document}

\title{
Invasion, polymorphic equilibria and fixation of a mutant social allele in group structured populations}
\author{ 
Roberto H. Schonmann$^{1}$,
Renato Vicente$^2$,
Robert Boyd$^{3}$}

\date{\today}

\maketitle

\begin{center}
1. Dept. of Mathematics, University of California at Los Angeles,
CA 90095, USA \\
2. Dept. of Applied Mathematics, Instituto de Matem\'atica e Estat{\'\i}stica,
Universidade de S\~ao Paulo, 05508-090, S\~ao Paulo-SP, Brazil\\
3. Dept. of Anthropology, University of California at Los Angeles,
CA 90095, USA
\end{center}





\begin{abstract}  
Stable mixtures of cooperators and defectors are often seen in nature. 
This 
fact is at odds with predictions based on linear public goods games
under weak selection. That model implies fixation either of cooperators or of defectors,
and the former scenario requires a level of group relatedness larger than the cost/benefit ratio, 
being therefore expected only if there is either kin recognition or 
a very low cost/benefit ratio, or
else under stringent conditions with low gene flow.  
This motivated us to study here  
social evolution in a large class of group structured populations,
with arbitrary multi-individual interactions among group members
and random migration among groups.  
Under the assumption of weak selection, 
we analyze the equilibria and their stability. 
For some significant models of social evolution 
with non-linear fitness functions, 
including contingent behavior in iterated public goods games and threshold models,
we show that three regimes occur, depending on the 
migration rate among groups. 
For sufficiently high migration rates, a rare cooperative allele A cannot invade a monomorphic 
population of asocial alleles N. 
For sufficiently low values of the migration rate, allele A can invade the population, 
when rare, and then fixate, eliminating N.
For intermediate values of the migration rate, allele A can invade 
the population, when rare, producing a polymorphic equilibrium, in which it 
coexists with N. 
The equilibria and their stability 
do not depend on the details of the population structure. 
The levels of migration (gene flow) and group relatedness
that allow for invasion of the cooperative allele
leading to polymorphic equilibria with the non-cooperative allele
are common in nature. 
\end{abstract}





\vspace{2mm}
\noindent{\bf Emails: \ }{\href{mailto:rhs@math.ucla.edu}{rhs@math.ucla.edu}, \ \
\href{mailto:rvicente@ime.usp.br}{rvicente@ime.usp.br}, \ \
\href{mailto:rboyd@anthro.ucla.edu}{rboyd@anthro.ucla.edu}








\

\section{Introduction}
\label{sec:intro}

In \cite{SVC} we studied evolution of behavior in group structured populations,
in a population structure that 
we called {\it two-level Fisher-Wright framework with selection and migration} (2lFW), 
which incorporates intra- and inter- group competition
and migration. 
The methods used in that paper allow 
for the study of arbitrary group interactions, 
that affect the reproductive success of group members as  
linear or non-linear functions of the number of group member exhibiting each possible behavior.
Those methods allow for the analysis under an arbitrary strength of selection, 
but are limited to the problem of invasion of the population by a rare mutant allele.
Here we add the assumption that selection is weak, and we obtain a characterization 
of all the equilibria and their stabilities in the same framework, as well as in a large 
class of biologically relevant generalizations. This allows us to 
identify three regimes, determined by demographic parameters of the model. 
(1) When the migration rate among groups is sufficiently high, 
invasion by a rare mutant social allele is not possible. 
(2) When the migration rate is sufficiently low, 
invasion happens and leads to fixation. 
(3) For intermediate values of the migration rate, 
invasion happens and leads to a polymorphic equilibrium. For some significant models of social 
interaction (e.g., iterated linear public goods games with contingent cooperation \cite{Joshi, BR, SVC}, 
threshold payoff functions arising from instance from coordinated punishment \cite{AS, BGB}) 
all three regimes occur.

We will consider a range of population structures
that extend 2lFW in various natural ways.
In 2lFW groups compete directly and more efficient groups split and replace less efficient ones
(see \cite{SVC} for a detailed introduction, or Section \ref{sec:ps} for a brief introduction). 
In Section \ref{sec:ps} we extend the 2lFW population structure in four ways.  
In Gen.1 we will allow for a 
separation of the time scales in which groups compete and in which 
individuals reproduce. 
In Gen.2 we will assume that  
groups may or not replace other groups, but more efficient groups will become larger than 
the less efficient ones, and contribute more migrants to the population, allowing cooperative 
alleles to spread. 
In Gen.3 we will allow for arbitrary patterns of recolonization, for occasional formation
of new groups, and for differential migration rates depending on phenotype, group size and group composition. 
And in Gen.4 we will allow for more general reproductive systems than the Fisher-Wright one.  

Importantly, our characterization of the equilibria and their stability
(Sections \ref{sec:ws} and \ref{sec:nm})
is essentially the same for all the population structures 
that are considered, suggesting a substantial level of generality. In particular, 
the conditions for invasion of a rare allele, obtained in \cite{SVC} are extended to this 
large class of population structures, when selection is weak.  
This is 
important because in \cite{SVC} these conditions were used to show that 
cooperative and even strongly altruistic alleles can invade a group structured population 
under realistically modest levels of group relatedness (realistically high levels of gene flow; 
as reported, e.g., in 
\cite{HC} (Tables 6.4 and 6.5), 
\cite{MHam} (Table 4.9) and \cite{BG}),
produced simply by viscosity, without any kin-recognition mechanism. 
Particularly noteworthy are two facts. 
First, conditions for invasion do not depend on the relation between the time scales in which individuals 
reproduce and groups compete (Gen.1). Group competition can, for instance, occur in the form of occasional direct 
conflict, which does not need to take place often (in the time scale of the individuals generations) 
for selection to favor cooperative alleles. 
This 
follows from the fact that 
carriers of cooperative 
allelles only incur reproductive costs when cooperation is needed, i.e., at times of conflict, 
in this example. So that the time scales in which costs and benefits are relevant are both the time scale
of group competion and therefore are the same.      
Second, even when each group
creates a single group in the next generation, so that groups  
do not compete directly, 
and with regulation of the population occuring only at the group level (Gen.2), the 
conditions for invasion are not changed, and invasion can occur under the same realistic conditions. 
This is due to the elasticity in the 
groups size, that allows for groups with more cooperators to be slightly larger. Since we are only 
considering weak selection, that enlargement effect will be small and hard to observe in practice. 
  
The tools that we provide here for computing equilibria/stability are easy to implement.
And when groups are large they simplify into expressions that only depend 
on the scaled gene flow parameter, or alternatively on the standard group relatedness parameter.  
In this way we provide an easy way to analyze 
evolution of social behavior under weak selection in group structured 
populations. As we explained in \cite{SBV}, one of the most commonly used approaches to that problem, 
the Taylor-Frank method \cite{TF, Taylor, WGF, GWW}, assumes implicitly that marginal fitness functions are 
linear functions of the number of mutants in a group. 
In other words, the reproductive success of group members has to be described in good approximation by a 
linear public goods game. 
The same remark applies to any methodology  
in which one expresses the fitness of 
a focal individual in terms of partial derivatives with respect to the focal
individual's phenotype and the phenotype of the individuals with whom the focal interacts,
as in \cite{Rou, LKWR, LR}. 
Our approach allows instead for arbitrary marginal fitness functions, that typically
result from complex interactions involving several individuals
\cite{Joshi, BR, KJC-B, Aviles, HMND, GWW, Vee1, AS, 
GYO, CRL, SDZ,  
NTW1,
HP, BJRB,
BG, BGB}. 
This feature of our approach also distinguishes it from and makes it much more applicable than 
approaches based on interactions between pairs of individuals (diadic interactions) 
\cite{Que2, LK}.   

Our results help explain the common occurrence of polymorphic equilibria between cooperators 
and defectors in nature 
\cite{HP}, \cite{GT}, \cite{R-GGWG}, \cite{GYO}.
As pointed out in \cite{R-GGWG}, when the reproductive success of group members 
is modeled as a linear public goods game and selection is weak, invasion leads to 
fixation, without the possibility of polymorphic equilibria. That paper explored 
strong selection as an alternative way of obtaining polymorphic equilibria. 
Our results show that the issue is not necessarily one of strong versus weak selection,
but that the non-linear nature of the interaction in the groups may yield those equilibria even when 
selection is weak.  

\section{Population structures}
\label{sec:ps}
We begin with a brief reprise of the 2lFW framework (see \cite{SVC} for further motivation and details).  
In 2lFW haploid individuals live in a large number $g$ of groups of size $n$,
and are of two genetically determined phenotypic types, A or N.  
Generations do not overlap,
reproduction is asexual and the type is inherited by the offspring.
(Except when stated otherwise, we suppose that mutations do not occur within the time scales 
being considered.) 
The relative fitness ($w$) of a type A, and that of a type N, 
in a group that has $k$ types A, are, respectively,
$w^A_k = 1 + \delta v^A_k$ and $w^N_k = 1 + \delta v^N_k$,
with the convention that $v^N_0 = 0$, i.e., $w^N_0 = 1$. 
The quantities $v^A_k$ and $v^N_k$ represent life-cycle payoffs  
derived from 
behavior, physiology, 
etc. 
The parameter $\delta \geq 0$ 
indicates the strength of selection.
When $\delta > 0$ is small, selection is weak, 
and we will also refer to the payoff functions  
$v^A_k$ and $v^N_k$ in this case as marginal fitness functions.
The creation of a new generation in the 2lFW through  
inter- and intra- group competition, 
followed by migration at rate $m$, 
is described as follows.
{\it Fisher-Wright intergroup competition:}
Each group in the new generation  
independently descends from
a group in the previous generation,
with probabilities proportional to group average fitness
$\bar w_k = \frac{k w^A_k + (n-k)w^N_k}{n}$. 
{\it Fisher-Wright Intragroup competition:}
If a group 
descends from  
a group with $k$ types A, then it will have $i$ types
A with probability $P_{k,i} 
= \mbox{bin}(i\, | \, n,k w^A_k / n \bar w_k)$,
where the binomial probability $\mbox{bin}(i|n,q)$ is the 
probability of $i$ successes in $n$ independent trials, each with 
probability $q$ of success. 
{\it Migration:}
Once the new $g$ groups have been formed according to the  
two-level  
competition process, a random fraction $m$ of the individuals migrates.
Migrants are randomly shuffled.  
{\it Note:}  
The assignment of relative fitness to the groups in
the fashion done above 
is a necessary and sufficient condition \cite{KG-S} for individuals in the 
parental generation to have each
an expected number of offspring proportional to their personal relative fitness. 


We denote by $f_k(t)$, $k =0, ..., n$, the 
fraction of groups in generation $t$ that have exactly $k$ types A.
And we denote by $p(t) = \sum_{k=1}^n (k/n) f_k(t)$ the frequency of types A in the population.   
Because the number of groups is large, by the law of large numbers, $(f_0(t), \dots, f_n(t))$ evolves as a 
deterministic (non-linear) dynamical system in dimension $n+1$.
Our main concern in this paper is in finding the stable and unstable equilibria for this dynamical system. 
This will be done under the assumption that selection is weak, i.e., $\delta << 1$.  

\

2lFW will be generalized in several ways in this paper.
These generalizations are motivated by 
biological considerations and, under weak selection, create little additional mathematical difficulties. 
We motivate and introduce the generalizations in Gen.1 - Gen.4 below. 
We then summarize 
the assumptions that we require, in a more abstract fashion, at the end of the section. 

\

\noindent{\bf (Gen.1) Separation of generation time and competition time scales:} 
In 2lFW, each group produces in the next generation a number of offspring groups that 
has a Poisson distribution. This may not be realistic in several biological applications, 
and therefore we will make the following more general assumption. 
Each group in generation $t$ produces in the next generation a random 
number of groups with a given distribution (that does not depend on the number of groups $g$) and has 
mean proportional to its average group fitness ($\bar{w_k}$ if it has 
$k$ types A and $n-k$ types N). 
The numbers of groups created in generation $t+1$ by the $g$ groups in generation $t$ are independent 
random variables, conditioned on the total number of groups in generation $t+1$ to be again $g$.
These assumptions allow one to use the law of large numbers as above, and
also in Section \ref{sec:drift} below. This extension of the framework
allows for important applications, in which groups typically create
a single group in the next generation, but occasional extinctions and recolonizations allow also 
for a group to create 0 or 2 groups. This generalization allows for a separation of the
time scales in which individuals reproduce (1 generation) and the possibly much longer time
scale in which groups create new groups. 
The kinds of cooperative behaviors that we have in mind here are those that only take place 
when there is direct group competition (e.g., inter-group conflict, or competition among groups 
for space opened by another group's  extinction). 
In this way the inter-generation time scale is not the same as the scale of cooperation 
and competition within and among the groups. Those time scales of cooperation and competition
are the same and are given by the 
time scale of group reproduction/extinction. 
In the analysis in this paper, we never have to consider 
what the distribution of the number of groups created by a group is
(only the mean of this distribution matters), so that this generalization
incurs no mathematical cost at all. (This is also the case of all the results in \cite{SVC}, 
including the results under strong selection, which therefore extend with no change to the 
population structures defined by Gen.1.) 

\

\noindent{\bf (Gen.2) Variable group size, local regulation:}
To motivate a second generalization of the 2lFW, consider the special case of the generalization
introduced in the previous paragraph, in which each group creates exactly one group in the  
next generation. 
For instance, each group might be located in a different habitat. In this case the population 
structure is the well known infinite island structure.  
Now, each group has size $n$ and generates exactly one group of size $n$. This implies that 
the average absolute fitness (expected number of offspring) of members of each group
must be 1, and therefore each group must have the same relative fitness. 
But this is incompatible with 
the behavior 
of types A benefiting (in terms of reproductive success) other members of their group at a cost 
(in terms of reproductive success) to themselves that is lower than the benefit produced. 
In such situations, the average fitness of the members of groups with many cooperators must be larger than 
that of groups with few cooperators (cooperative and altruistic behavior should be defined as ``benefiting the group'',
or more precisely, benefiting the average member of the group, in reproductive terms). 
In conclusion: there is no room for cooperative/altruistic behavior in this population 
structure, as it stands. 
This is the setting of \cite{Taylor}, which is often referred to for this conclusion. 
By assuming a fixed number 
$n$ of members in each group, we force $\bar{w}_k$ to be independent of $k$. 
Biological populations often admit elasticities in population and group size 
that are not incorporated in this formulation \cite{Taylor+}. 
But because we will only consider weak selection ($\delta << 1$) in this paper, we 
can easily extend the framework to accomodate this sort of elasticity, without having to be 
concerned with the details, and without affecting the analysis. To see this observe two facts.
First, the assumption 
that all the groups have size $n$ in 2lFW is just a mathematical idealization/simplification of a population
in which there is a typical group size close to $n$ most of the time. 
Second, all the fitnesses 
are different from 1 by a term of order $\delta$.
Keeping these facts in mind, we generalize now the 
2lFW framework, by allowing group size to be different from $n$, but only with probability of
order $\delta$. 
For this purpose, we suppose that in each generation, each group is created with a 
random size that has a distribution that 
depends on the size and composition of its parental group
in an appropriate way. 
If the parental group has size $n$, then the size of its offspring groups 
concentrates almost all the probability on the size $n$, except for 
a probability of order $\delta$. 
For other sizes of the parental group, we suppose that its offspring groups have 
positive probability (even if $\delta = 0$) of having size $n$. 
Under these conditions in each generation all but a fraction of order $\delta$ of groups 
will have size $n$. And following the history of any single group, it will 
have size $n$ almost all the time, the exception being a fraction of generations of order $\delta$.
Intuitively, $n$ is the typical group size, but ocasional deviations are allowed, 
where ``ocasional'' is quantified by $\delta$. 
Now, since the fitnesses $\bar{w}_k$ only differ from 1 by a term of order $\delta$, 
this new framework can accomodate any such fitness function, meaning that the 
expected size of a group whose parent group has $k$ types A and $n-k$ types N will 
be precisely $\bar{w}_k$.  
(Even if the framework only allows for group sizes $n-1$, $n$ and $n+1$,
this is sufficient flexibility to allow for any fitness functions $w^A_k$, $w^N_k$ 
of the kind that we are considering.
This intuitive idea is checked in detail in Appendix A, where further biological 
considerations are also discussed.)   
Note that when $\delta = 0$, as in Section \ref{sec:drift}, groups will always have size $n$,
so that this generalization is of no consequence in that section.  
When $\delta > 0$ is small this is not the case, but typically groups still have size $n$.
And because the case of small $\delta > 0$ can be analyzed as a small perturbation of the 
case with $\delta = 0$ (separation of time scales), again the generalization in this paragraph
will be of no consequence in Sections \ref{sec:ws} and \ref{sec:nm}, 
where weak selection will be analyzed in 
this way.   

\

\noindent{\bf (Gen.3) Exceptional groups, 
exceptional migrants,  
weakly variable migration rates, 
variable number of groups:}
When groups vary in size, the migration patterns in large and small groups can 
differ from the typical ones. 
As in Gen.2 above
we can incorporate this generalization into the mathematical analysis
provided that except for probabilities of order $\delta$, groups will have size $n$ and 
behave regularly. 
For the same reason, a fraction of order $\delta$ of migrants may behave in 
a way that is different from the regular one, even when they come from groups of 
size $n$.
There is no restriction on the migration patterns of the 
exceptional migrants. These considerations allow for instance for the inclusion in the framework
of groups that temporarily have size 0, and may represent an empty habitat. 
Such an empty habitat may, for instance, be colonized by migrants from the migrant population,
without having a parental group, or alternatively be colonized by a single 
parental group. The recolonization scheme will not affect our analysis and results,
since in each generation only a fraction of order $\delta$ of the groups may have 
been recolonized in this generation or descend from a group that was recolonized
recently. In other words, if we select a group at random in some generation, it is 
likely (except for probability of order $\delta$) to come from a lineage of regular 
groups deep into the past.     
Similarly, we can allow some migrants to form a completely new group (which 
then has no parental group), but we suppose that in each generation, only a 
fraction of order $\delta$ of migrants behaves in this way. 
Moreover, all the individuals may migrate at rates that depend on their type and that of the other
members of their group, but for groups of size $n$ this variability must be 
within bounds of order $\delta$. This means that the migration rate of an 
individual of type $*$ (A, or N), in a group with a total of $k$ types A and 
$n-k$ types N can be of the form $m(1+\delta u^*_k)$, where $u^*_k$ are arbitrary.    
Some of the assumptions above imply
that the number of groups is not constant, and it is natural to assume that 
it is regulated by ecological forces towards the typical number $g$. This 
idea can be incorporated into the mathematical framework, by assuming that, if
in generation $t$ the number of groups is $g_t$, then the 
fitness of the groups is multiplied by a regulation factor $H(g_t/g)$, where 
$H(1) = 1$
and $-1 < dH(s)/ds < 0$. (These assumptions on $H$ imply that when the number of groups is smaller than 
$g$ it tends to grow and when it is larger than $g$ it tends to shrink, and that it 
does not tend to evershoot the target value $g$.) 
Under this assumption, there is no 
need to condition (as we did in Gen.1) on the total number of groups being created
in a given generation. We instead simply suppose that each group creates a number of groups
with a distribution that only depends on its group fitness (with the factor $H(g_t/g)$,
common to all the groups).

\

\noindent{\bf (Gen.4) General intragroup reproduction system:}
Biological considerations may 
require the distribution of types of members of a group to depend on the distribution of the
types of the members of the parental group in a way that is different from that given by the 
Fisher-Wright 
sampling $P_{k,i} = \mbox{bin}(i|n,k w^A_k/n\bar{w}_k)$. 
We can assume an arbitrary intragroup transition matrix $P_{k,i}$, 
under the biologically meaningful conditions that when $\delta = 0$, 
$\sum_{i} i P_{k,i} = k$ (neutral drift) and 
$P_{0,0} = P_{n,n} = 1$ (no mutations).
Generalizing the population structure 
in this fashion produces some minor but cumbersome complications in 
(\ref{TjkFW}), in Section \ref{sec:drift}. For simplicity we will only compute the more
general expression in Appendix B. 
Other than this, all the statements in Section \ref{sec:drift} (no selection: $\delta = 0$) 
and Section \ref{sec:ws} (weak selection: small $\delta > 0$) hold for a 
general intragroup transition matrix $P_{k,i}$. 
Additional, and again biologically meaningful, conditions on $P_{k,i}$ will be needed for 
the results in Section \ref{sec:nm} (case of large group size $n$ and small migration rate $m$) to hold.
For simplicity   
we will restrict ourselves to Fisher-Wright intragroup sampling 
when providing numerical examples in Section \ref{sec:examples}. But as we will 
explain in Section \ref{sec:nm} and Appendix B, when $n$ is large, 
these numerical results should also be good approximations
under broad conditions on the intragroup sampling scheme.  

\



In the following sections the setting is always that of 2lFW possibly generalized by Gen.1, Gen.2 and Gen.3.
Everything will also apply under Gen.4, unless stated otherwise, in which case the appropriate 
modifications needed are presented in Appendix B. 
Even additional generalizations would be covered by the results, provided that 
they satisfied the assumptions stated explicitly below.   
It may be surprising that so many different aspects of the population structure will produce almost no
difference in the analysis and the results. This is a feature of weak selection. 
The key fact
is that the reference population structure with $\delta = 0$ is the same and that it can be easily
analysed. The evolution under a small $\delta > 0$ is then a perturbation to this reference system, 
and to first order in $\delta$ does not depend then on the details of the perturbation. 
This makes the concept of weak selection a very powerful tool in shedding light on evolutionary 
patterns. (In contrast, the analysis under strong selection is highly dependent on the population 
structure.)  

The assumptions 
used in Section \ref{sec:drift} are: 

\

{\noindent}
{\bf Assumptions about the reference population structure ($\delta = 0$):}
(1) From  
generation $t$ to generation $t+1$, each group produces  
independently (possibly conditioned on the total number of groups produced being $g_{t+1} = g$) 
a random number of groups with a common distribution with mean $H(g_t/g)$, where $H$ is
a regulation function that 
has $H(1) = 1$ and has derivative $-1 < dH(s)/ds <0$ (as in Gen.3).  
(Note that if we condition on $g_{t} = g$, in each generation,  
then each group must produce in the average one group in the next generation.) 
(2) Each group being created has $n$ individuals, and the number $i$ of individuals of type A in this group, 
conditioned on the number $k$ of individuals of type A in its parental group, is given by 
a transition matrix $P(k,i)$ that satisfies $\sum_{i} i P_{k,i} = k$ and $P_{0,0} = P_{n,n} = 1$
(see Gen.4 for motivation). 
(The case of Fisher-Wright sampling is defined by $P(k,i) = \mbox{bin}(i|n,k/n)$.)  
(3) Once a new generation is created, migration occurs at rate $m$, with migrants being randomly shuffled. 

\

The assumptions that are required for the perturbation approach used in Section \ref{sec:ws}
are also simple to state: 

\ 

\noindent{\bf Assumptions on weak selection (of strength $\delta$):}
(1) Suppose that a focal individual is selected at random in a given generation, 
and we observe how many individuals of type A and of type N are in the focal's group. We assume that 
except for probabilities of order $\delta$ there is no difference in taking this sample 
from the actual population or from the reference population with $\delta = 0$. 
(Formally, we are requiring the distance between the two distributions in total variation
to be at most of order $\delta$. This means that the probability of finding a given composition of the 
focal's group, when computed using the actual population or the reference population, only differ by an
amount of order $\delta$.) 
(This is precisely what we observed to hold in Gen.2 and Gen.3).  
(2) The average number of offspring of a randomly chosen focal individual who is in a group 
with a total of $k$ types A and $n-k$ types N is proportional to $w^*_k = 1 + \delta v^*_k$, 
where * is the type, A or N, of the focal individual. (Note that we do not need to make any
assumption about these relative fitnesses when the size of the focal's group is different 
from the typical size $n$.)   

\

The separation of the conditions in two sets, one defining the restrictions on the $\delta = 0$ 
population structure, and one specifying how ``$\delta$ effects '' are allowed to affect it, 
is a central element of our analysis. The idea of separation of time scales associated to 
weak selection, that we will exploit in Section \ref{sec:ws} is a standard one 
\cite{Rou, LKWR, RR, Lessard}, 
but in the literature one often assumes a population structure (Wright's infinite island structure 
is a very common choice) that is not affected (even slightly) when $\delta$ is positive.
(For instance this is the case in \cite{Taylor}, and as we explained in Gen.2, 
for this reason the setting of that paper rules out the possibility of altruistic behavior.)
But weak selection can be the result of a weak force that affects the reproductive 
success of individuals through its slight effect on the population structure (e.g., rare 
extinction of groups due to natural disasters or warfare). This idea
is incorporated in the very mild restrictions that are being imposed on the effects
of order $\delta$. The complete flexibility in the kinds of $\delta$ effects that can modify 
the population structure allows for a large range of natural and biological processes to
be described in a concise simplified way.     

Should one generalize the population structure further? Two of the constraints on the reference 
($\delta = 0$) population structure that will be used in Section \ref{sec:drift} are the fixed
size of groups, and the completely random unstructured migration at rate $m$ (each individual
once born is a migrant with probability $m$ independently of anything else, and migrants are 
randomly shuffled). Releasing these constraints to allow say for groups of other sizes in the
reference population, or structured migration according to some migration matrix, would 
require important modifications in Section \ref{sec:drift}. Such modifications are amenable to
some extent to the methods of population genetics, but substantially more eleborate than the 
case treated here. It is to avoid such technical issues that, up to effects of order $\delta$, 
we restrict ourselves here to 
groups of fixed size and to unstructured migration. These are 
typical idealizations of population with groups sizes that fluctuate around some typical value,
and have migration ranges that are not very short.    

\section{Neutral drift}
\label{sec:drift}
Before studying the evolution of $p(t)$ under weak selection, we consider the 
case in which selection is completely absent, i.e., $\delta = 0$. In this case,
there is no bias towards types A or N. Hence, 
each individual in generation $t$ is of type A with probability $p(t-1)$, and
again by the law of large numbers, $p(t)=p(t-1)$.
We can therefore regard $p(t) = p$ as constant in the following computation.  
When a new generation is being born, 
each individual in a group has its type, independently, 
with probability $1-m$ sampled from the parental group of its group, or, with probability $m$ sampled from the 
metapopulation, where types A occur with frequency $p$.
Therefore, 
assuming that the intragroup transition matrix is given by Fisher-Wright sampling (see Appendix B for 
the general case), 
the probability $f_k(t)$ that a group selected at random in generation $t$ will have
$k$ types A will satisfy the recursion 
\be
f_k(t) \ = \ \sum_j T_{j,k}\, f_j(t-1),
\label{recursionforf}
\ee
where 
\be
T_{j,k} \ = \ \mbox{bin}(k|n, (1-m)(j/n)  + mp).
\label{TjkFW}
\ee
This means that 
$(f_0(t), ... , f_n(t))$ 
evolves as the probability distribution of a Markov chain, 
with transition matrix $T_{j,k}$ 
and converges to
its stationary distribution that we denote by 
$\varphi(p) = (\varphi_0(p), ... , \varphi_n(p))$, given by the stationarity condition
\be
\varphi_k(p) \ = \ \sum_j 
T_{j,k}
\, \varphi_j(p),
\label{stationarity}
\ee
and $\sum_k \varphi_k(p) = 1$.
(In the case of the infinite islands population structure, the results above can be found in
\cite{Wak}, p.413.) 
Note that 
\be
\sum_k k \varphi_k(p) = n p,
\label{mean}
\ee
since the left-hand-side is the expected number of types A in a randomly chosen group,  
there are $n$ individuals in each group, and
each individual is type A with probability $p$. 

Some extreme cases will play important roles later in the analysis. 
(See Appendix C, %
for an alternative explanation/derivation of the claims below.)
Note first that
(\ref{mean}) implies that 
\begin{eqnarray}
\varphi(p) 
& \to & (1,0,...,0,0), \ \ \ \ \mbox{as} \ \ \ p \to 0,
\label{varphiaspto0}
\\
\varphi(p)
& \to & (0,0,...,0,1), \ \ \  \ \mbox{as} \ \ \ p \to 1.
\label{varphiaspto1}
\end{eqnarray}
The case $m=1$ clearly has 
\be
\varphi_k(p) = \mbox{bin}(k|n,p).
\label{varphim1}
\ee
In the opposite extreme, when $m \to 0$, the irreducible transition matrix $T_{j,k}$ 
converges to a transition matrix with traps at $0$ and $n$, so that 
\be
\varphi(p)
\to (1-p,0,...,0,p), \ \ \  \mbox{as} \ \ \ m \to 0,
\label{varphiasmto0}
\ee 
where we used (\ref{mean}) again.

In each one of the equilibria, $\varphi(p)$, 
the average 
relatedness $R$ between distinct members of a randomly chosen group takes the same value
(Appendix D provides a short review of this and the other well known claims in this paragraph).
To express it in a way that 
does not require Fisher-Wright intra-group reproduction 
(includes Gen.4),  we recall that the imbreeding effective 
group size, $n_{\mbox{eff}}$, is defined as the inverse of the probability that, when 
a new group is created and before migration takes place, two randomly chosen members
of this group have the same mother. In the case of Fisher-Wright intragroup sampling, we then  
simply have $n_{\mbox{eff}} = n$. With this definition, 
\be
R \ = \ \frac{(1-m)^2}{n_{\mbox{eff}} - (n_{\mbox{eff}}-1)(1-m)^2} 
\ \approx \ \frac{1}{1+2mn_{\mbox{eff}}},
\label{R}
\ee 
where the approximation is good when $m$ is small. Here relatedness can be defined through lineages 
(identity by descent -- IBD), regression coefficients, or Wright's $F_{ST}$ statistics. 

\section{Weak selection}
\label{sec:ws}
When selection is weak, i.e., $0 < \delta << 1$, a well known separation of time scales argument 
\cite{Rou, LKWR, RR, Lessard}
applies in the following way.  
Starting with $p(0) = p$ the population reaches the quasi-equilibrium with distribution $\varphi_k(p)$
(up to an error term of order $\delta$)
in a time of order $1/m$, without feeling the effect of selection. But then selection produces changes 
in $p(t)$ at a rate of order $\delta$. 
This is so because the fitnesses of types A and types N differ by an
amount of order $\delta$.
To implement this idea, 
we write $\Delta p(t) = p(t+1) - p(t)$ and recall the well known formula 
\be
\bar W \Delta p \ = \ p (W^A - \bar W) \ = \ p(1-p) (W^A - W^N),
\label{DeltapWAWN}
\ee
where $W^A$ 
and $W^N$ are the average fitnesses of types A and N, 
and $\bar W = p W^A + (1-p) W^N$ is the average fitness of all individuals
(all in generation $t$, that is being omited in the notation, for simplicity).
To compute $W^A$ and $W^N$, we choose at random a focal individual from the population.
We denote by $I_{\bullet}$ a random variable that takes the value 1 if the focal is type A and 0 
if the focal is type N. And we denote by $w_{\bullet}$ a random variable that equals the relative fitness
of the focal individual. Then $W^A = \mbox{E}(w_{\bullet}|I_{\bullet}=1)$ and 
$W^N = \mbox{E}(w_{\bullet}|I_{\bullet}=0)$. To compute
these conditional expectations, we also denote by $K$ the random variable that gives the
total number of types A in the focal's group. Then
\be
\mbox{Pr}(K=k|I_{\bullet}=1) \ = \ \frac{\mbox{Pr}(I_{\bullet}=1, K=k)}{p} \ = \
\frac{\mbox{Pr}(I_{\bullet}=1|K=k) f_k(t)}{p} \ = \ \frac{k f_k(t)}{np}.  
\label{Bayesian}
\ee
This means that if the focal is type A, it will have probability $k f_k(t)/np$
of being in a group with exactly $k$ types A.
(The factor $k$, that tilts the distribution $f_k(t)$,
can be easily understood as a bias term. Given that the focal is
type A, it is more like that it was chosen from a group with many types A). 
An analogous computation shows that if the focal is type N,  
it will have probability $(n-k) f_k(t) / n(1-p)$
of being in a group with exactly $k$ types A. 
Therefore, $W^A = \sum_k k f_k(t) w^A_k / np$, and 
$W^N = \sum_k (n-k) f_k(t) w^N_k / n(1-p)$. 
When $f_k(t) = \varphi_k(p)$, we have then   
$W^A = 1 + \delta V^A(p)$ and $W^N = 1 + \delta V^N(p)$, where 
\begin{eqnarray}
V^A(p) \ & = & \ 
\frac{\sum_{k} k \varphi_{k}(p) v^A_{k}}{np},
\label{VA}
\\
V^N(p) \ & = & \ 
\frac{\sum_{k} (n-k) \varphi_{k}(p) v^N_{k}}{n(1-p)}.
\label{VN}
\end{eqnarray}
And now from (\ref{DeltapWAWN}), 
\be
\Delta p \ = \ 
\delta p(1-p) [V^A(p) - V^N(p)] \ + \ O(\delta^2).
\label{Deltap}
\ee
This equation tells us that if we accelerate time by a factor $1/\delta$
(so that the rescalled time is $s = t \delta$), then $p(t)$ will evolve as the 
solution $x = x(s)$ to the differential equation 
$dx/ds = x(1-x)[V^A(x)-V^N(x)]$,
started from $x(0) = p$. 
The equilibria are the values of $p$ for which 
\be
D(p) \ = \
p(1-p) [V^A(p) - V^N(p)] \ = \ 0.
\label{equilibria}
\ee 
They are $p=0$, $p = 1$ and the internal ones, where $V^A(p) = V^N(p)$. 
Their stability is also readily obtained, by observing the sign of $D(p)$ close to each one of them
(decreasing $D(p)$  $\, \Rightarrow \, $ stability, 
\ increasing $D(p)$ $\, \Rightarrow \, $ instability).
Stability of the dynamical system can be equivalently thought off as stability with respect to a 
small perturbation in the frequency of alleles, or with respect to a small but positive mutation rate
\cite{KM} (see also SOM of \cite{SVC}, Section 9).      

In \cite{SVC}, we were concerned with the stability of the equilibrium with $p=0$, i.e., we were 
interested in determining when rare mutants A can invade a population of types N. 
This problem was solved there (for 2lFW) with no assumption on the strength of selection. 
The simplification of that general solution in the case of weak selection was then provided 
in display (2) in that paper. (We review the content of that condition from \cite{SVC} in Appendix E, for use in 
Section \ref{sec:examples}.) 
That condition is therefore equivalent to 
the condition that $\lim_{p \to 0} [V^A(p) - V^N(p)] > 0$. But from (\ref{varphiaspto0}), we have 
$\lim_{p \to 0} V^N(p) = v^N_0 = 0$, and the condition for invasion by rare types A becomes
\be
\lim_{p \to 0} \, V^A(p) \ = \
\sum_{k=1}^n \, \left ( \lim_{p \to 0} \frac{k \varphi_k(p)}{np} 
\right) v^A_k \ > \ 0.
\label{invasion}
\ee 

The case in which $m=1$, in which groups assort randomly in each generation,
is equivalent to the trait-group framework (see, e.g., Sec. 2.3.2 of \cite{Okasha}). In this case 
(\ref{varphim1}) and 
simple manipulations with binomial coefficients transform (\ref{VA}), 
(\ref{VN}) and (\ref{Deltap}) into the well known forms
\be
V^A(p) \  =  \ \sum_{k=1}^n \, \mbox{bin}(k-1|n-1,p) \, v^A_{k},  \ \ \ \ \
V^N(p) \ = \ 
\sum_{k=1}^n \, \mbox{bin}(k-1|n-1,p) \, v^N_{k-1},
\label{VANm=1}
\ee
\be
\Delta p \ = \ \delta p(1-p) 
\sum_{k=1}^n \, \mbox{bin}(k-1|n-1,p) \, (v^A_k - v^N_{k-1})
 \ + \ O(\delta^2).
\label{Deltapm=1}
\ee
And the condition (\ref{invasion}), for invasion by rare types A, becomes simply $v^A_1 > 0$.
Under the strong altruism condition $v^{A}_{k} < v^N_{k-1}$
(meaning that each type A would be better off mutating into a type N), 
(\ref{Deltapm=1}) implies the well know result \cite{MJ, KG-SF} that, when $m=1$, the only
stable equilibrium is the one with no types A, and types N fixate when present. 
The same is therefore also true, under this strong altruism condition, when $m$ is close to 1. 

In the opposite extreme, 
when $m \to 0$, 
using (\ref{varphiasmto0}) we transform 
(\ref{VA}), 
(\ref{VN}) and (\ref{Deltap}) into 
\be
V^A(p) \  =  \ v^A_{n},  \ \ \ \ \
V^N(p) \ = \  v^N_{0} \ = \ 0,
\label{VANmto0}
\ee
\be
\Delta p \ = \ \delta p(1-p) 
\, v^A_n
 \ + \ O(\delta^2).
\label{Deltapmto0}
\ee
This means that if $v^A_n > 0$, then for $m$ close to 0, not only will rare types A invade the
population, but they will fixate. The only equilibria are then the ones with $p=0$ and $p=1$, 
the former being unstable and the latter stable. 

Consider now a model in which the following two conditions hold:
$$
\mbox{(C1)} \ \ v^A_1 < 0, 
\ \ \ \ \ \ 
\ \ \ \ \ \ 
\ \ \ \ \ \
\mbox{(C2)} \ \ v^A_n > 0.
$$
These conditions are associated, for instance, to a social allele A, that promotes a certain behavior
that it costly to an isolated actor, but that also provides a net benefit to each actor, when all members of the
group behave in this way.  In this case, from the conclusions in the last two paragraphs, 
we learn that there are two critical values of the migration parameter $m$, namely $0 < m_f \leq m_s < 1$,
which play the following roles. 
The critical point $m_s$, already introduced in \cite{SVC}, is  the threshold value for types A to
be able to invade when rare 
(for $m > m_s$ allele A cannot invade when rare).
The critical value $m_f$ is the threshold value for types A to fixate after invasion
(for $m < m_f$ allele A not only invades 
when rare, but it then fixates). (The subscripts `s' and `f' stand for `survival' and `fixation', 
respectively, of the allele A.) 

Because (\ref{R}) establishes a one-to-one relationship between $m$ and $R$, and because it is 
common to measure $m$ indirectly through $R$, it is natural to also introduce $R_s$ and $R_f$ 
as the values of $R$ that correspond to $m_s$ and $m_f$, respectively. 

In Section \ref{sec:examples}, we will see important examples for which $m_f < m_s$ ($R_s < R_f$), 
so that there is 
an intermediate regime in which types A invade and evolve to a polymorphic equilibrium. 
This is nevertheless not the case, as is well known \cite{Ham75,Rou,R-GGWG}, for the linear public goods game, 
in which at a cost $C$ to itself each type A contributes a benefit $B$ shared by the 
other members of its group. 
As we review in Appendix F, 
for this model $V^A(p) - V^N(p) = -C + BR$ does not depend on $p$,
so that $R_s = R_f$ is the solution to Hamilton's equation $C = BR$.   

\section{Limit of large group size and small migration rate}
\label{sec:nm}
We turn now to the case in which $n$ is large and $m$ is small. It is a classical result of S. Wright
\cite{Wright31, CK, Wak} 
(regarding generalization Gen.4, the conditions on $P_{k,i}$ for this result to hold are discussed
in Appendix B -- basically one only has to assume that individuals 
produce statistically independent numbers of offpring according to their fitnesses, conditioned on the 
appropriate total number of offspring in the group) 
that if we take the limit in which 
\be
n \to \infty, \ \ \ \ \ m \to 0, \ \ \ \ \ 2mn_{\mbox{eff}} \to  l,
\ \ \ \ \ \left(\mbox{so that, in particular, $R \to \frac{1}{1+l}$}\right),
\label{limit}
\ee 
then for every $0 < y < 1$, 
\be
\sum_{k \leq yn} \varphi_k(p) \to \int_0^y \mbox{beta}(x|lp,l(1-p)) dx,
\label{limbeta}
\ee
where 
$\, \mbox{beta}(x|\alpha,\beta) \, $ 
is the density of a beta distribution with parameters $\alpha$ and $\beta$, 
which up to a normalization constant is given by $x^{\alpha -1} (1-x)^{\beta -1}$. This means that when we 
rescale $k$ by dividing it by $n$, producing a variable $x = k/n$ in the interval $[0,1]$, the 
distribution of $x$ given by $\varphi_k(p)$ is close to that of a $\mbox{beta}(x|lp,l(1-p))$. 
Suppose we also have the approximations $v^A_k \approx \widetilde{v}^A_{k/n}$ and $v^N_k \approx \widetilde{v}^N_{k/n}$, 
(uniformly in $k$) for some piecewise continuous functions $\widetilde{v}^A_x$ and $\widetilde{v}^N_x$, $0 < x < 1$.
Then, for large $n$, (\ref{VA}) and (\ref{VN}) can be replaced by 
\be
V^A(p) \ = \ \int_0^1 \, \mbox{beta}(x|lp + 1, l(1-p)) \, \widetilde v^A_x \, dx, 
\label{VAlimit}
\ee
\be
V^N(p) \ = \ \int_0^1 \, \mbox{beta}(x|lp , l(1-p) + 1) \, \widetilde v^N_x \, dx. 
\label{VNlimit}
\ee
The changes in the values of the parameters $\alpha$ and $\beta$ in the beta 
distributions that appear in (\ref{VAlimit}) and (\ref{VNlimit}), from those of the  
beta distribution in (\ref{limbeta}) are due to the bias terms $k/n$, or $(n-k)/n$,
in (\ref{VA}) and (\ref{VN}), respectively. These yield factors 
$x$ and $(1-x)$, respectively, which are incorporated into the density of the beta
distribution, modifying its parameters. Note that the factors $p$ and $1-p$ in the 
denominators in (\ref{VA}) and (\ref{VN}) are simply normalizing factors, and  are
absorbed into the normalization of the beta distributions.

The condition (\ref{invasion}) for invasion by rare mutants A becomes in the 
limit (\ref{limit}), 
\be 
\lim_{p \to 0} \, V^A(p) \ = \
\int_0^1 \, \mbox{beta}(x|1,l) \, \widetilde{v}^A_x \, dx
\ > \ 0,
\label{invasionlimit}
\ee  
or, equivalently,
\be 
\int_0^1 \, (1-x)^{l-1} \, \widetilde{v}^A_x \, dx
\ > \ 0,
\label{invasionlimit+}
\ee  
which has already appeared implicitly in display (3) in \cite{SVC}, where it was derived in 
a different way.   

It is important to emphasize that the results in this section 
do not depend on 
the details of the population structure, including the reproductive system. Only the scaled 
population parameter $m n_{\mbox{eff}} = l/2$, that represents gene flow, 
or, equivalently, 
the average group relatedness, $R = 1/(1+2 m n_{\mbox{eff}})$,
appears.
In the case of the 
linear public goods game, $\widetilde{v}^A_x = -C + Bx$, and
(\ref{invasionlimit+}) is equivalent to Hamilton's rule $C < BR$. As we pointed out in 
\cite{SVC}, it is natural to consider (\ref{invasionlimit+}) as an extension of that rule
to non-linear games. 
It is pleasant to observe how this combines two major ideas in 
evolutionary biology. The idea from population genetics of finding scaling parameters 
that summarize the relevant features of a large class of population structures. And  
Hamilton's key idea, that genetic relatedness could provide, under appropriate assumptions,
the mathematical conditions for 
the proliferation of a cooperative allele. While both ideas have their limitations, the
results in this section provide a case in which both are vindicated. 

\section{Examples}
\label{sec:examples}
The conditions for equilibrium, stability and invasion, derived in the previous sections apply 
with equal ease to any marginal fitness functions $v^A_n$, $v^N_k$. We focus in this section 
on two examples that are representative of some important biological mechanisms. 

In the first of these
two examples (IPG), non-linearities in the marginal fitness functions $v^A_k$ and $v^N_k$, result from 
the iteration of a basic game and contingency of behavior in each iteration. By assuming that the
underlying base game is a linear public goods game (PG), one can analyze the effect of non-linearity 
resulting only from the iterative/conditional aspect of the life-cycle interaction among group 
members. This also makes the PG a special (extreme) case of the class of models being considered
(the case when the number of iterations is 1). 

In the second example (THR) we consider a very simple type of non-linearity in  
the marginal fitness functions $v^A_k$ and $v^N_k$, that incorporates synergetic 
effects of cooperation, by requiring a minimum number of cooperators for successeful 
outcomes, as well as saturation effects, when there are many cooperators. We then also 
consider iterations of such a game and observe the effect of the number of iterations. 

The examples in this section illustrate some of our main concerns and conclusions. Non-linear 
marginal fitness functions associated to iteration and contingencies 
are natural in group structured populations, where group members 
live together for spans of time that are long compared to the time frame of their 
vital activities. For instance, many vertebrates live together and hunt together 
hundreds or thousands of times during their life-cycle. This gives plausibility to 
figures like the ones in the graphs below. For reasonable values of the parameters 
(group size, cost/benefit ratio for each iteration, number of iterations), we obtain 
often values of $R_s$ that are modest (below $10 \%$) 
as compared to available data on $F_{ST}$ values
\cite{HC} (Tables 6.4 and 6.5), 
\cite{MHam} (Table 4.9) and \cite{BG} (see Appendix D for the relation between relatedness 
and $F_{ST}$).
In contrast, the values of $R_f$ are substantially higher. This indicates that 
coexistence of cooperators and defectors should not be surprising. 

The iterative nature of the games and the conditional behavior of the cooperators
play central roles in allowing for low values of $R_s$. One can only expect 
a modest value of $R_s$, if the benefits of cooperation in  
groups with many cooperators are substantially 
larger than the costs of cooperation in groups with few cooperators. This may be unlikely 
to happen with a single 
iteration of a biologically realistic game, due to physical and biological constraints.
Also if a game is iterated, but behavior does not change from iteration to iteration, 
the same negative conclusion applies, since payoffs are then simply multiplied by the number of 
iterations.  
But once nature has endowed individuals with mechanisms that prompt them to change 
behavior based on past experience, the goal becomes realistic, as we see below.       

\

\noindent
{\bf (IPG) Iterated public goods game. Types A cooperate conditionally:} 
\begin{eqnarray*}
v^A_k & = & \left\{
                \begin{array}{cc}
                                    -C + (k-1)B/(n-1),  & \mbox{if \  $k \leq a$},  \\
                                   T \, (-C + (k-1)B/(n-1)), & \mbox{if \  $k > a$}, 
                \end{array}
\right.  \\
v^N_k & = &  \left\{
                \begin{array}{cc} 
                                    kB/(n-1),  & \mbox{if \  $k \leq a$},  \\
                                    T \, kB/(n-1), & \mbox{if \  $k > a$},  
                \end{array}
\right.
\end{eqnarray*}
for constants $0 < C < B$, $T \geq 1$ and $a \in \{1,2,...,n-1\}$.
This example was first studied independently in \cite{Joshi, BR}, basically in the 
context in which $m = 1$.  
In this example a public goods game (PG) is repeated a random number of times $\tau \geq 1$, 
with average $\E(\tau) = T$.  Each time each member of the 
group can cooperate at a cost $C$ to itself, resulting in a benefit $B/(n-1)$ to each
one of the other members of its group. 
Defectors incur no costs and produce no benefits.
We suppose that types A cooperate in the 
first round, and afterwards only cooperate if at least $a$ other members of the 
group cooperated in the previous round. This is often called a ``trigger stategy'' (with
threshold $a$) and is a generalization of the well known
tit-for-tat strategy, which corresponds to the case $n=2$, $a=1$.
In \cite{SVC} we focused on the case in which  
each type A cooperates in the first round, and in later rounds cooperates 
if and only if its payoff in the previous round was non-negative.  
This corresponds to 
$a$ taking the smallest integer value that is larger than or equal to $(n-1)C/B$
(since the PG has $v^A_k < 0$ if and only if $k < 1 + (n-1)C/B$).
In this important case, we call the strategy of types A
``payoff dependent contingent cooperation''. The idea is that after an act of
cooperation, each individual who cooperated receives a feedback, that indicates 
if cooperation should continue or not. If a negative value 
of $v^A_k$ is associated with a negative feedback, then types A 
will discontinue cooperation precisely according the
payoff dependent contingent cooperation rule. Types A in 
this example do not have to keep track of the identity of group member who cooperated 
or defected, and do not have to count how many cooperated. They only have to be predisposed 
to discontinue behaviors that hurt them (in net terms), and continue  
behaviors that are beneficial to them (in net terms).
The conditionally cooperative behavior of types A in this example 
is in this sense closely related to generalized reciprocity mechanisms \cite{RT}
with low cognitive requirements. 
The assumption that individuals discontinue behavior after a single unsuccessful participation 
is a simplification. When this is not a realistic assumption, one can  
interpret the parameter $T$ as the ratio between the typical 
number of repetitions of the activity and the typical number of unsuccessful attempts before 
cooperation is discontinued by a type A.


\begin{figure}[ht]
  \includegraphics[width=0.8\textwidth]{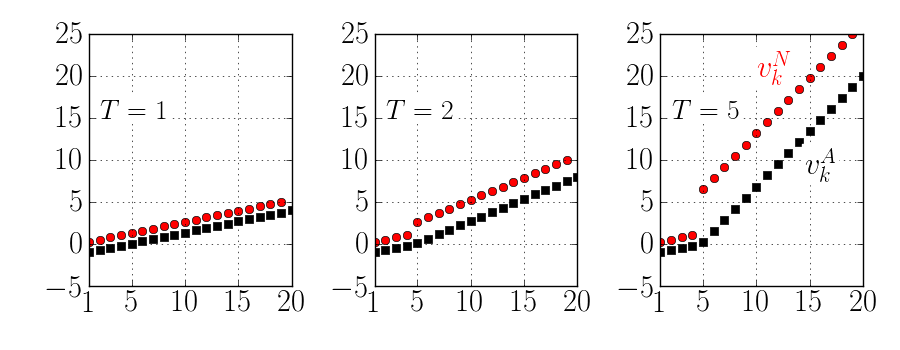}
  \caption{Payoff (marginal fitness) profiles for IPG. Payoffs $v^A_k$ for types A are represented by black squares, 
while red circles depict payoffs $v^N_k$ for types N. In these pictures, 
$n=20$, $C=1$, $B=5$, and $a=4$. From left: $T=1$, $T=2$ and $T=5$. }
\end{figure}
    
\

\noindent  {\bf (THR) Threshold model:}
\begin{eqnarray*}
v^A_k & = & \left\{
                \begin{array}{lc}
                                    -C,  & \mbox{if \  $k < \theta$},  \\
                                    -C + A, & \mbox{if \  $k \geq \theta$},
                \end{array}
\right.  \\
v^N_k & = &  \left\{
                \begin{array}{lc}
                                     0,  & \mbox{if \  $k < \theta$},  \\
                                     A', & \mbox{if \  $k \geq \theta$},

                \end{array}
\right.
\end{eqnarray*}
for positive constants $C$, $A$ and $A'$, and an integer $\theta \in \{1,2,...,n\}$.
The idea here is simple: the allele A carries a cost, but allows its carriers 
to gain benefits if sufficiently many are in the group. 
Unless otherwise stated, we assume that $A > C$. 
Types N obtain
benefits also when types A do, but we allowed for the possibility that these benefits 
are different from those of the types A. These payoff functions can be seen as 
simplifications of more realistic ones, in which payoffs to types A initially grow 
slowly with $k$, then steeply, and then quickly saturate. For instance, 
hunting of large pray may require a minimum number of hunters, but above that 
threshold number there can be little benefit to adding more hunters. 
Also the model of 
coordinated punishment of \cite{BGB} provides an example of this sort. 
In \cite{AS} a class of models that bridge between the PG and the THR was
introduced and their relevance discussed. Similarly to the results of \cite{BR} and \cite{Joshi}
for the IPG, 
it was shown in \cite{AS} that the THR can have polymorphic equilibria, with types A and
N coexisting, when assortment is random ($m=1$ case, trait-group framework). 
In this case, as for the IPG, types A can nevertheless never proliferate
when $p$ is small (the equilibrium with no types A, $p=0$, is stable). 
In \cite{Vee1}, the case $n=3$ of the threshold model (called there ``stag hunt game'') 
was discussed in connection to the conceptual issue of the role of Hamilton's rule.

The THR behaves in a very simple way under iteration. Suppose that the game is repeated $T$ 
times over a life-cycle. And suppose that types A are ``payoff dependent conditional 
cooperatores'', meaning that a type A cooperates in the first round, and in later rounds cooperates 
if and only if its payoff in the previous round was positive. 
The result is that types A cooperate exactly once if $k < \theta$, and cooperate $T$ times if 
$k \geq \theta$. The total payoffs to types A and to types N over the life-cycle are then
those of a THR with the same value of $C$, with $A$ replaced by $T(A-C) + C$ and $A'$ replaced 
by $T A'$. Fig.~\ref{fig:ITHRvs} illustrate the case in which for the base game 
$n=10$, $\theta = 4$, 
$C=1$, $A = A' =2$, so that the game iterated $T$ times has the parameters as 
indicated in the figure caption. 

\begin{figure}[ht]
  \includegraphics[width=0.8\textwidth]{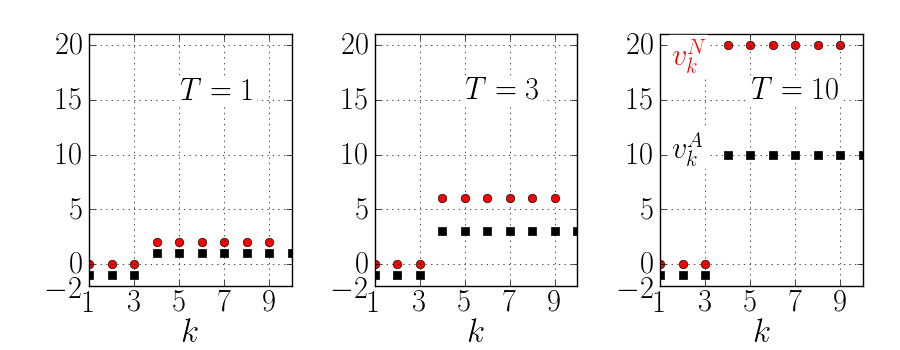}
  \caption{Payoff (marginal fitness) profiles for iterated THR. 
  Payoffs $v^A_k$ for types A are represented by black squares, 
  while red circles depict payoffs $v^N_k$ for types N. In these pictures 
  $n=10$, $\theta = 4$, $A = T + 1$ and $A' = 2T$. From left: $T=1$, $T=3$ and $T=10$. }
\label{fig:ITHRvs}
\end{figure}

\

\noindent 
{\bf Numerical analysis. Figures:}
We restricted ourselves to Fisher-Wright intra-group reproduction in the computations. 
So, strictly speaking, the figures below refer to 2lFW generalized by Gen.1, Gen.2 and Gen.3 
and not Gen.4. 
But, as explained in Section \ref{sec:nm} and Appendix B, when $n$ is large these results are
also good approximations to a broad class of intra-group reproductive systems addressed
in Gen.4, provided we set $n = n_{\mbox{eff}}$, as defined at the end of Section \ref{sec:drift}. 

In the figures below, the equilibria (indicated by the frequency $p$ of types A)
where obtained from solving (\ref{equilibria}), and their stability was obtained 
from the sign of $D(p)$ in their neighborhood (decreasing $D(p)$  $\, 
\Rightarrow \, $ stability, 
\ increasing $D(p)$ $\, \Rightarrow \, $ instability).   
Instead of the migration rate $m$, or the gene flow $n m$, we expressed 
the results as functions 
of the relatedness parameter $R$, given by (\ref{R}). 
This allows easier comparison with $F_{ST}$ data   
\cite{HC} (Tables 6.4 and 6.5), 
\cite{MHam} (Table 4.9) and \cite{BG} (see Appendix D for the relation between relatedness 
and $F_{ST}$). 
We computed $R_s$ independently of the computations of the equilibria, by the methods of 
\cite{SVC} (see Appendix E). A similar computation was made of the point $R_s^N$, above 
which types N cannot invade a population of types N 
(i.e., above which the point with $p=1$ is stable). 
These values are indicated with dashed vertical lines (red and magenta) on the graphs that
display the equilibria as functions of $R$. The agreement with the stability of the 
equilibria $p=0$ and $p=1$, obtained from (\ref{equilibria}) is clear in the pictures.
We are not aware of any general procedure for computing $R_f$, since it is defined in 
global terms (above it, the equilibrium with $p=1$ becomes the only stable equilibrium). 
But typically we have $R_f = \max \{R_s , R_s^N \}$, which happens in all our
pictures, and allows for the determination of $R_f$ in these cases.    
Most commonly we observed $R_f = R^N_s$, but in Fig.~\ref{fig:ITHRn=10,theta=8} we have an
exception.


\newpage

\vspace{1cm}

\begin{figure}[ht]
  \includegraphics[width=0.5\textwidth]{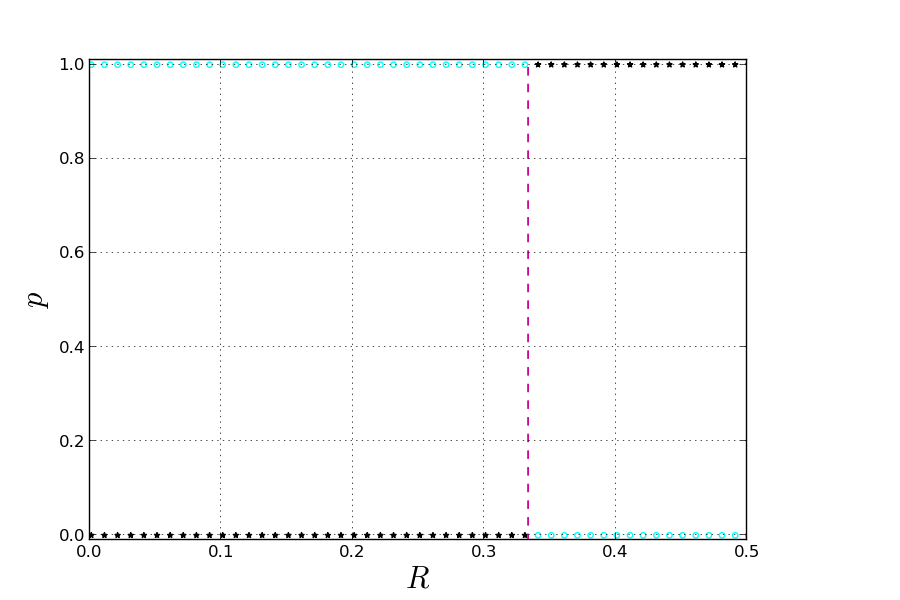} \hspace{-0.2cm} 
  \includegraphics[width=0.5\textwidth]{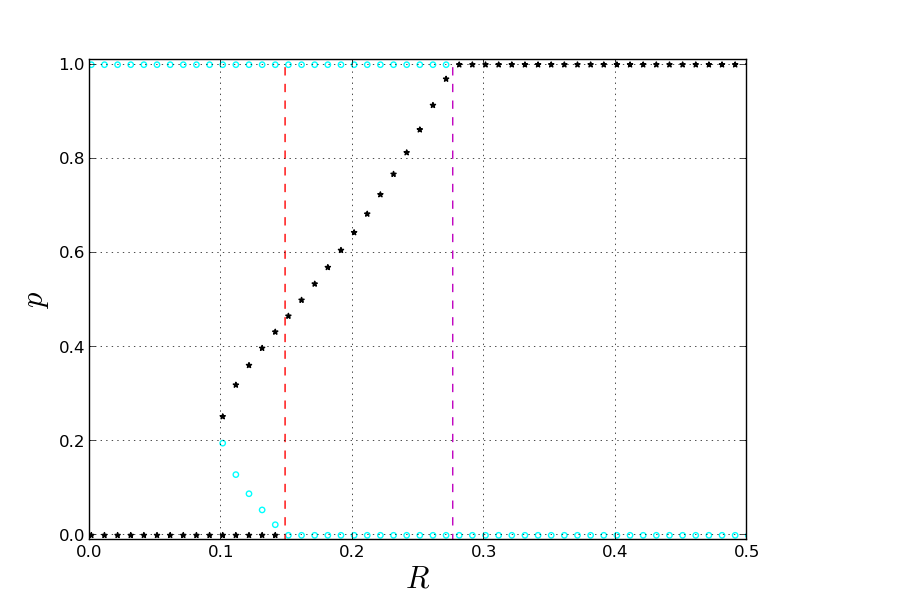} \vspace{0.2cm}
  \includegraphics[width=0.5\textwidth]{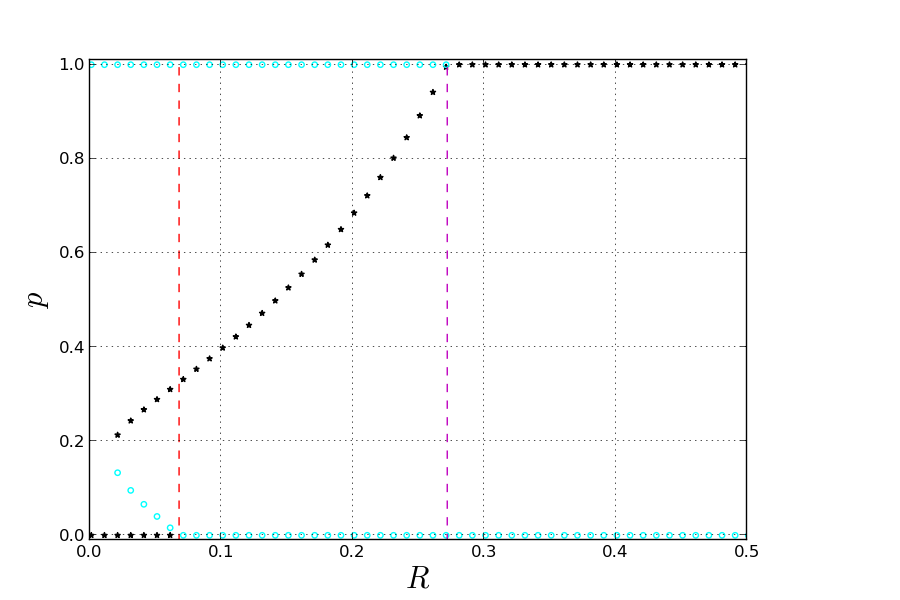} \hspace{-0.2cm} 
  \includegraphics[width=0.5\textwidth]{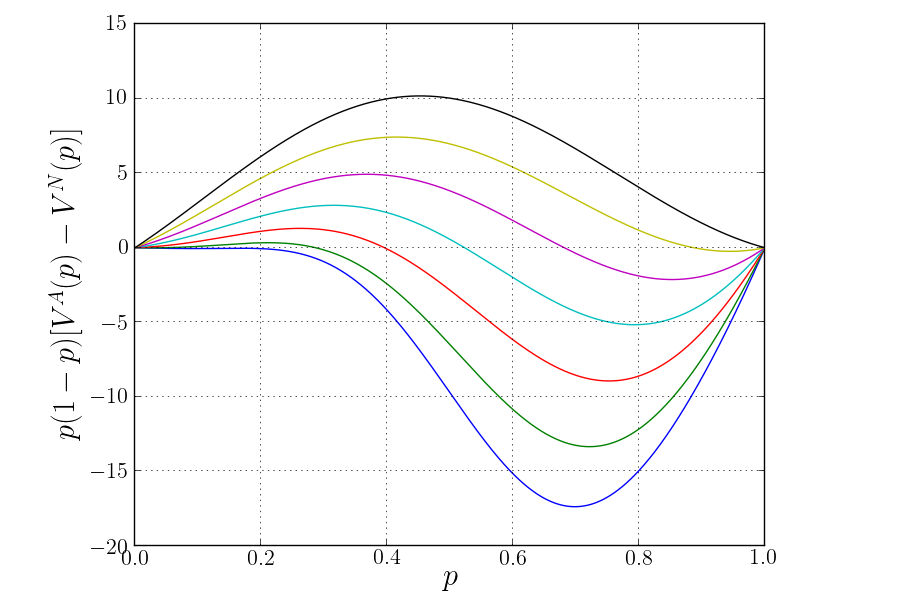} 
  \caption{{\bf IPG}: $n=10$, $C=1$, $B=3$, $a=4$. 
  ({\bf Top Left}) $\bm {T=1}$ ({\bf in this panel IPG = PG}): $R_s=R_f=33\%$ ($m_s=m_f=8.7\%$).
  ({\bf Top Right}) $\bm {T=10}$: $R_s=14.8\%$ ($m_s=20.3\%$) and  $R_f=27.6\%$ ($m_f=11\%$). 
  ({\bf Bottom Left}) $\bm {T=100}$: $R_s=6.8\%$ ($m_s=35\%$) and $R_f=27.2\%$ ($m_s=11.2\%$). 
  Open cian circles
  represent unstable fixed points, black stars represent stable fixed points. The critical relatedness for invasion
  by types A, $R_s$, is indicated by a red dashed line while the critical relatedness for invasion by types N, $R_s^N$, 
  is depicted as a magenta dashed line. The pictures show that here $R_f = R^N_s$.
  ({\bf Bottom Right}) Same parameter values from bottom-left panel: 
  profiles for $D(p) = p(1-p) [V^A(p) - V^N(p)]$ 
  at (from bottom to top) $R=1\%$ (blue), $5\%$ (green),$10\%$ (red),
  $15\%$ (cian), $20\%$ (magenta), $25\%$ (yellow) 
  and $30\%$ (black).
  }
\label{fig:IPGn=10}  
\end{figure}

\newpage

\begin{figure}[ht]
  \includegraphics[width=0.5\textwidth]{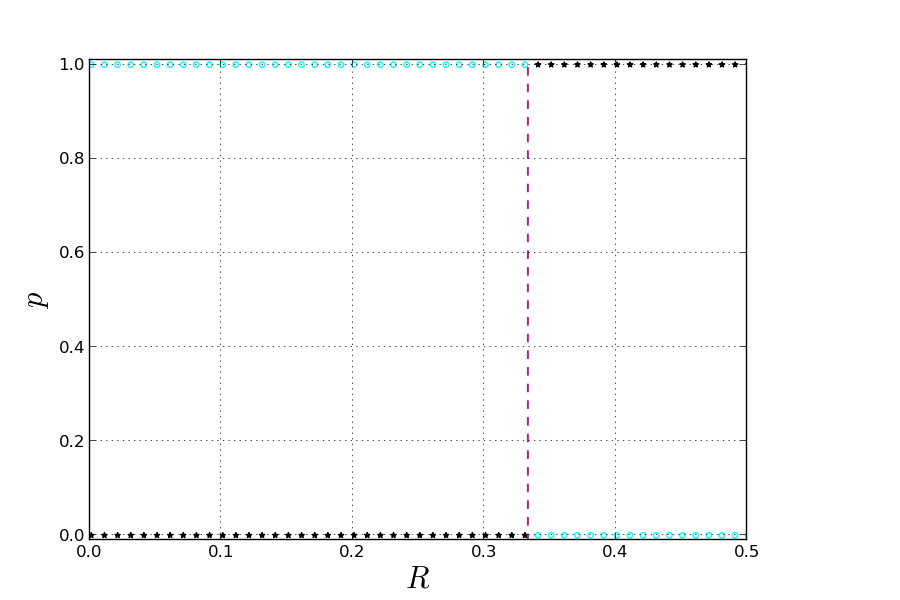} \hspace{-0.2cm} 
  \includegraphics[width=0.5\textwidth]{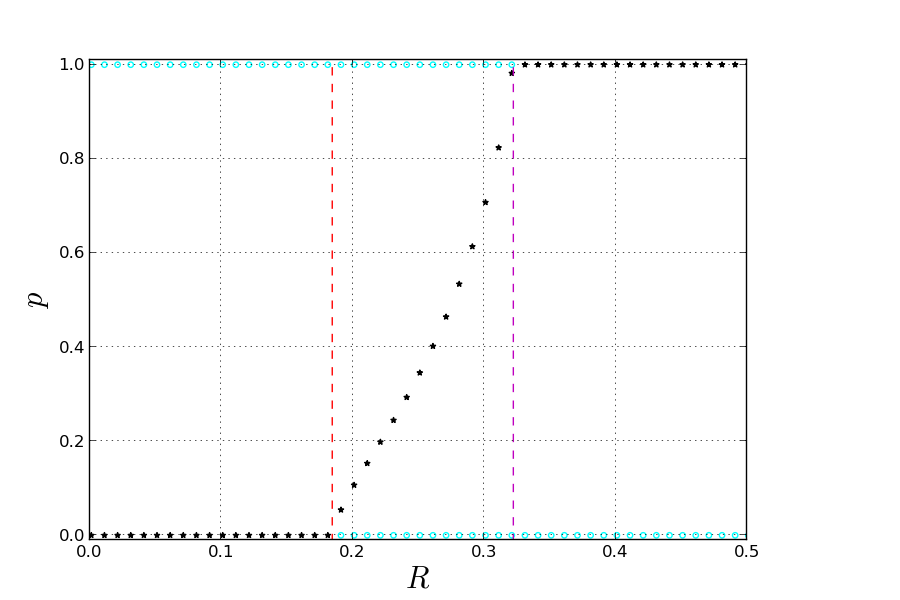} \vspace{0.2cm}
  \includegraphics[width=0.5\textwidth]{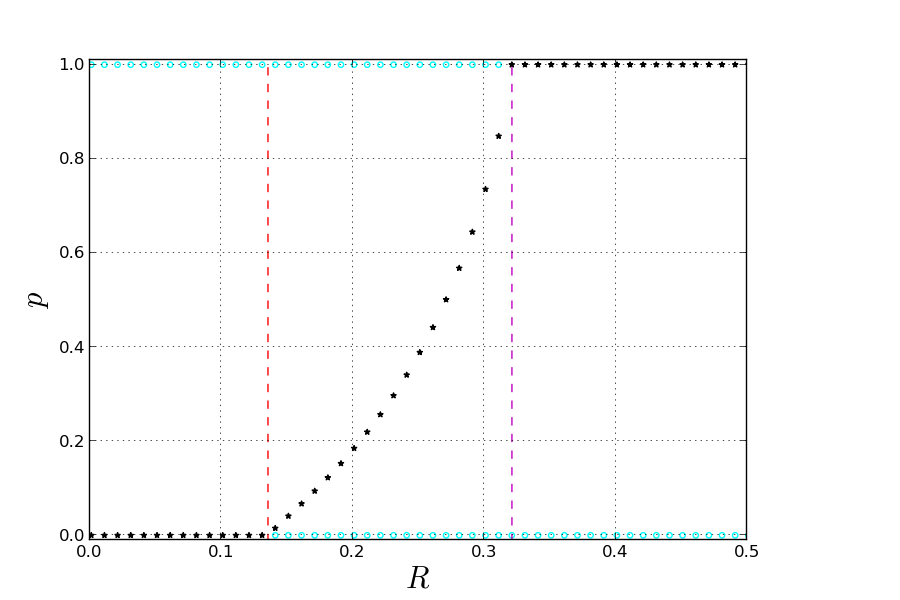} \hspace{-0.2cm} 
  \includegraphics[width=0.5\textwidth]{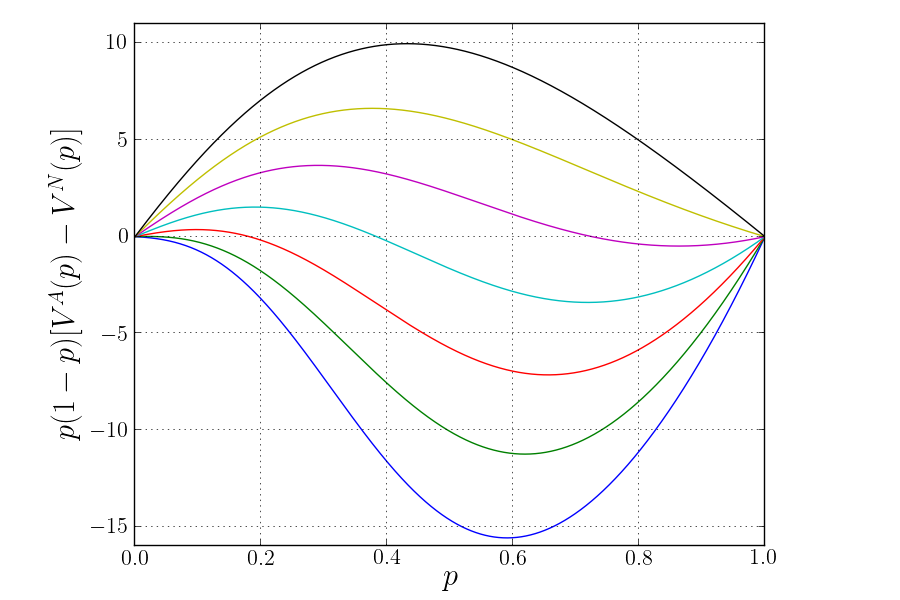} 
  \caption{{\bf IPG}: $n=18$, $C=1$, $B=3$, $a=4$. 
  ({\bf Top Left}) $\bm {T=1}$ ({\bf in this panel IPG = PG}): $R_s=R_f=33\%$ ($m_s=m_f=5.1\%$).
  ({\bf Top Right}) $\bm {T=10}$: $R_s=18.4\%$ ($m_s=10.4\%$) and  $R_f=32.2\%$ ($m_f=5.3\%$).
  ({\bf Bottom Left}) $\bm {T=100}$: $R_s=13.5\%$ ($m_s=14.1\%$) and $R_f=32.1\%$ ($m_s=5.4\%$). 
  Open cian circles
  represent unstable fixed points, black stars represent stable fixed points. The critical relatedness for invasion
  by types A, $R_s$, is indicated by a red dashed line while the critical relatedness for invasion by types N, $R_s^N$, 
  is depicted as a magenta dashed line. The pictures show that here $R_f = R^N_s$.
  ({\bf Bottom Right}) Same parameter values from bottom-left panel: 
  profiles for $D(p) = p(1-p) [V^A(p) - V^N(p)]$ 
  at (from bottom to top) 
  $R=10\%$ (blue), $15\%$ (green),$20\%$ (red),$25\%$ (cian), $30\%$ (magenta), $35\%$ (yellow) 
  and $40\%$ (black).
  }
\label{fig:IPGn=18}
\end{figure}

\newpage

\begin{figure}[ht!]
  \includegraphics[width=0.5\textwidth]{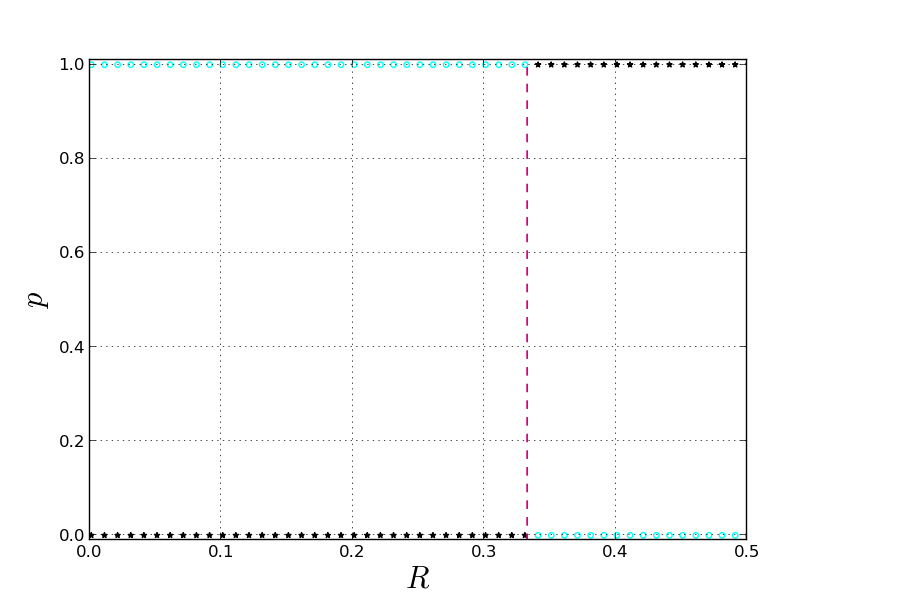} \hspace{-0.2cm} 
  \includegraphics[width=0.5\textwidth]{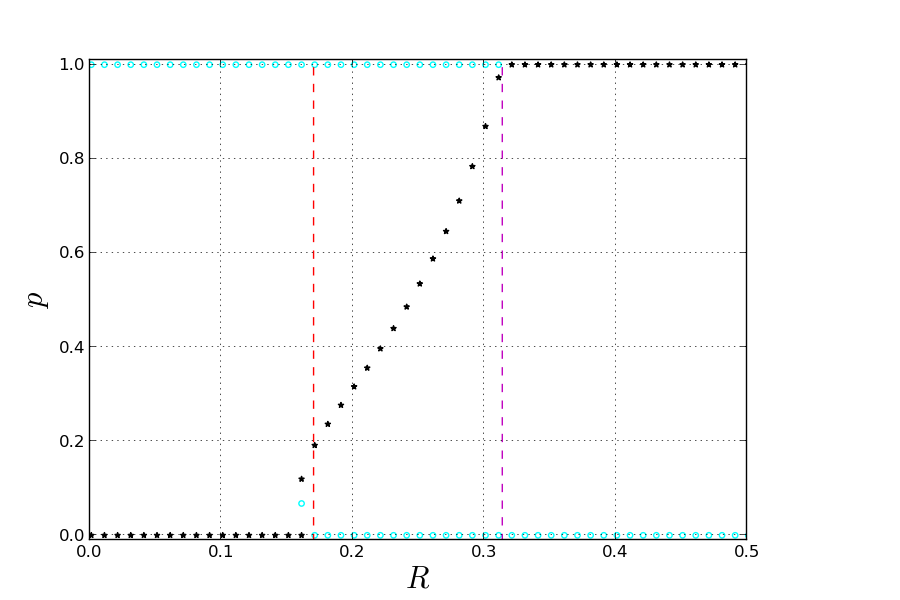} \vspace{0.2cm}
  \includegraphics[width=0.5\textwidth]{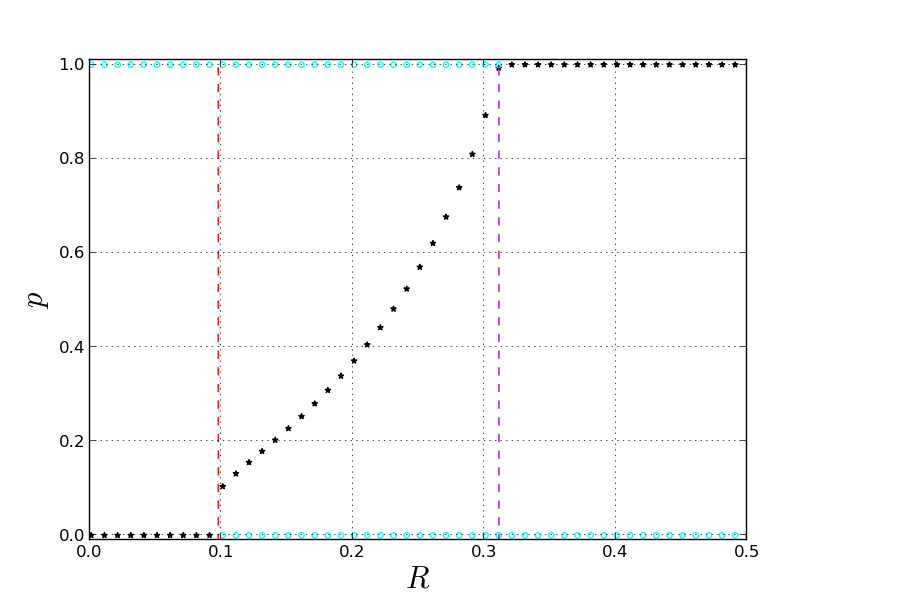}\hspace{-0.2cm}
  \includegraphics[width=0.5\textwidth]{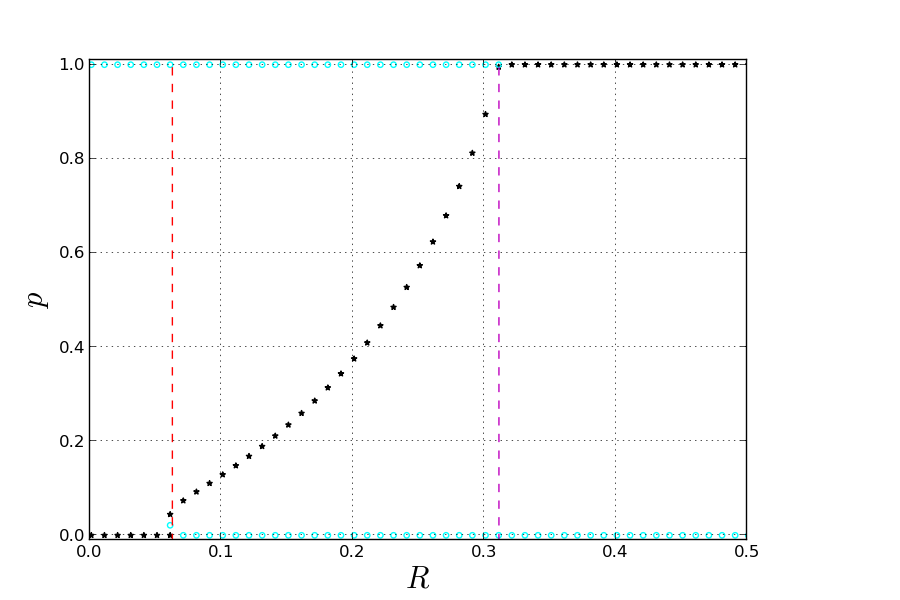} 
  \caption{{\bf IPG}: $n=50$, $C=1$, $B=3$, $a=16$. ({\bf Top Left}) $\bm {T=1}$ ({\bf in this panel IPG = PG}): 
  $R_s=R_f=33.3\%$ ($m_s=m_f=1.95\%$).
  ({\bf Top Right})   $\bm {T=10}$. $R_s=17\%$ ($m_s=4.55\%$) and  $R_f=31.4\%$ ($m_f=2.12\%$). 
  ({\bf Bottom Left}) $\bm{T=100}$: $R_s=9.76\%$ ($m_s=8.1\%$) and  $R_f=31.1\%$ ($m_f=2.14\%$).
  ({\bf Bottom Right}) $\bm{T=1000}$: $R_s=6.3\%$ ($m_s=12.3\%$) and  $R_f=31.1\%$ ($m_f=2.14\%$).  
  Open cian circles
  represent unstable fixed points, black stars represent stable fixed points. The critical relatedness for invasion
  by types A, $R_s$, is indicated by a red dashed line while the critical relatedness for invasion by types N, $R_s^N$, 
  is depicted as a magenta dashed line. The pictures show that here $R_f = R^N_s$.
  This model with these parameters are the same as in Fig.2 in \cite{SVC}, 
  which is restricted to the problem of invasion by types A, 
  but includes strong selection and comparison with the limit of large $n$. In this limit, 
  it follows from (\ref{invasionlimit+}), as computed in the supplementary material of 
  \cite{SVC}, that for large $T$ we have $R_s = -\ln(1-C/B)/\ln(T)$ approximately. Fig.2 in \cite{SVC}
  indicates that for $T = 10, 100, 1000$, this approximation is very good.
  As also observed in \cite{SVC}, in this example, types A are strongly altruistic, in the sense 
  that $v^A_{k} < v^N_{k-1}$
  (each type A would be better off mutating into a type N). 
  }
\end{figure}

\newpage


\newpage

\begin{figure}[ht!]
   \includegraphics[width=0.5\textwidth]{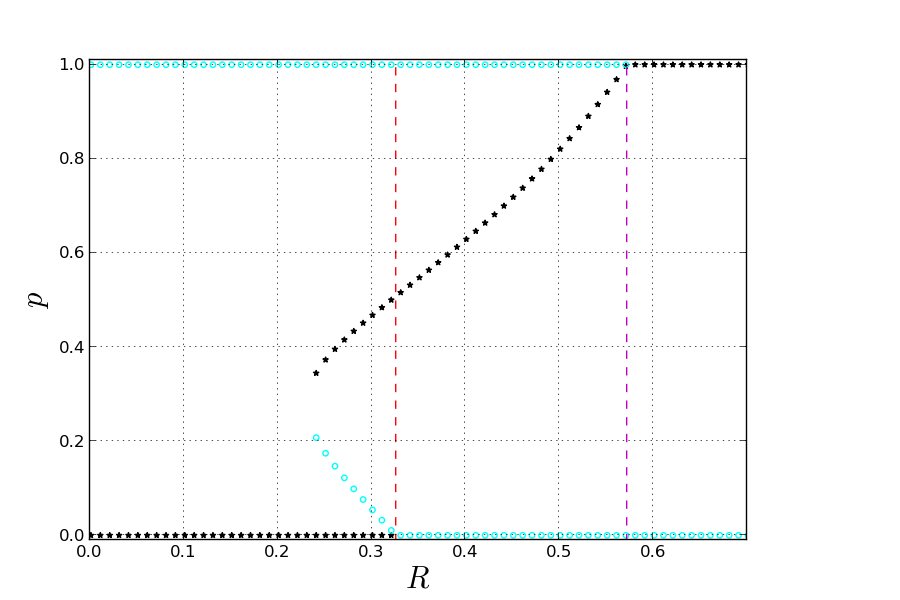} \hspace{-0.2cm} 
   \includegraphics[width=0.5\textwidth]{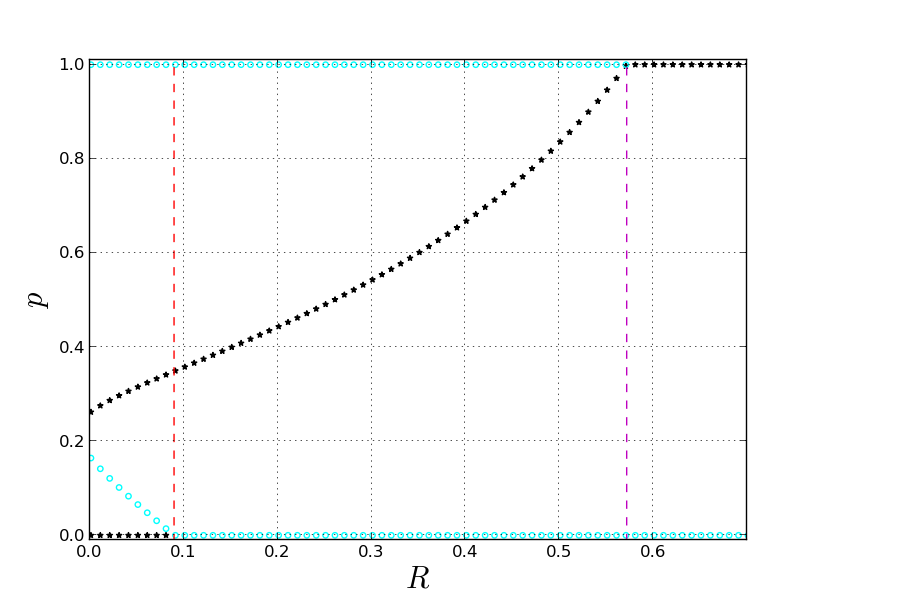}   
   \vspace{0.2cm}
   \includegraphics[width=0.5\textwidth]{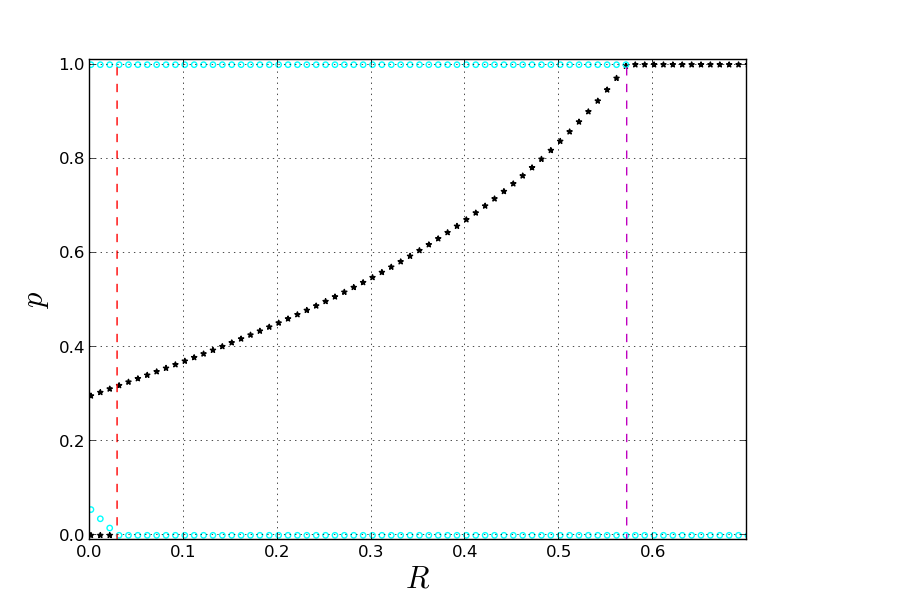} \hspace{-0.2cm} 
   \includegraphics[width=0.5\textwidth]{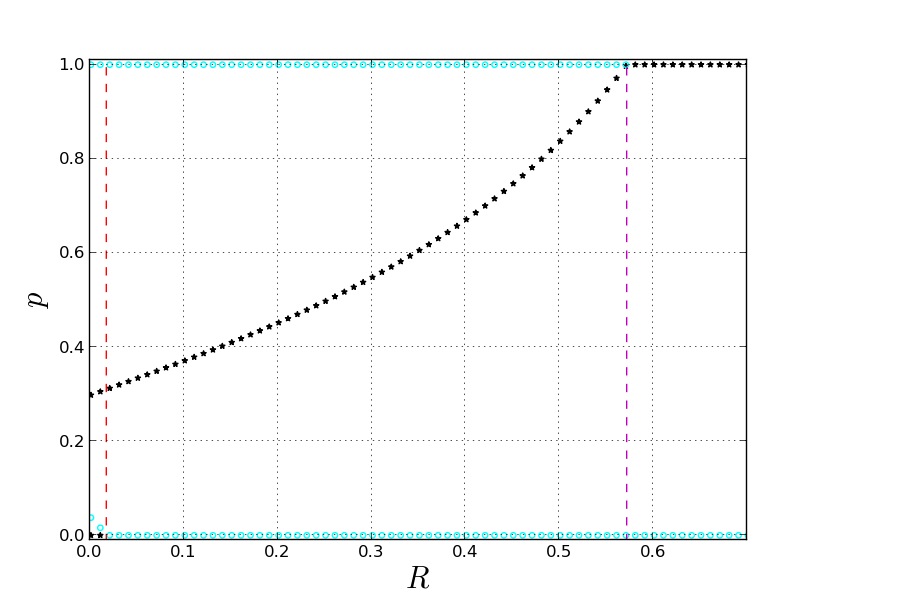}  
  \caption{{\bf THR}: $n=10$, $\theta = 4$, $C=1$, $A = T+1$, $A' = 2T$. 
  ({\bf Top Left}) $\bm {T=1} \ (\bm {A=2}, \, \bm{A^{\prime}=2})$: $R_s=32.6\%$ ($m_s=9\%$) and  $R_f=57.2\%$ ($m_f=3.5\%$). 
  ({\bf Top Right}) $\bm {T=10} \ (\bm {A=11}, \, \bm{A^{\prime}=20})$: $R_s=8.95\%$ ($m_s=29.6\%$) and  $R_f=57.2\%$ ($m_f=3.5\%$).
  ({\bf Bottom Left}) $\bm {T=100} \ (\bm {A=101}, \, \bm{A^{\prime}=200})$: 
  $R_s=2.9\%$ ($m_s=52.1\%$) and  $R_f=57.2\%$ ($m_f=3.5\%$). 
  ({\bf Bottom Right}) $\bm {T=300} \ (\bm {A=301}, \, \bm{A^{\prime}=600})$:
  $R_s=1.73\%$ ($m_s=61.2\%$) and  $R_f=57.2\%$ ($m_f=3.5\%$).
  Open cian circles
  represent unstable fixed points, black stars represent stable fixed points. The critical relatedness for invasion
  by types A, $R_s$, is indicated by a red dashed line while the critical relatedness for invasion by types N, $R_s^N$, 
  is depicted as a magenta dashed line. The pictures show that here $R_f = R^N_s$.
  }
\label{fig:ITHRn=10}
\end{figure}

\newpage

\begin{figure}[ht!]
   \includegraphics[width=0.5\textwidth]{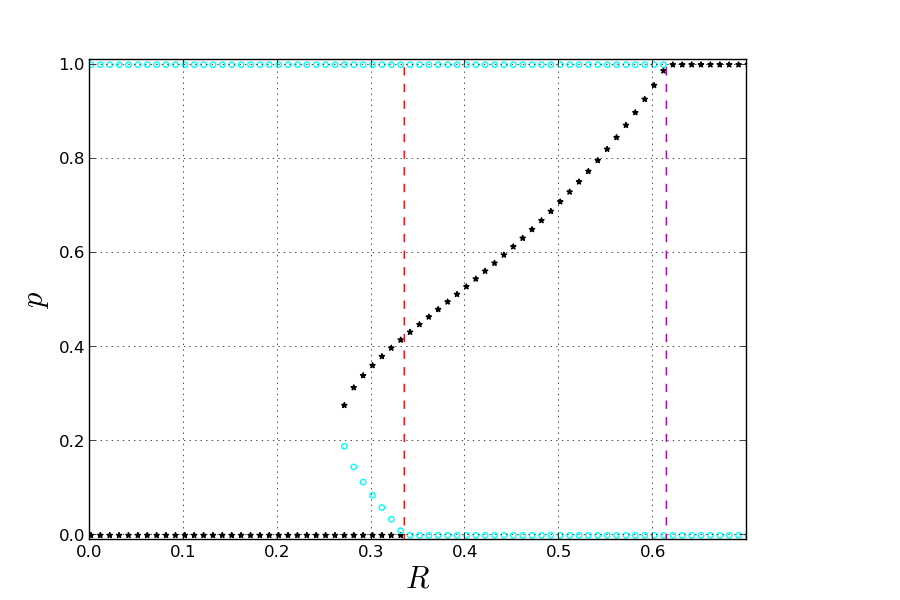} \hspace{-0.2cm} 
  \includegraphics[width=0.5\textwidth]{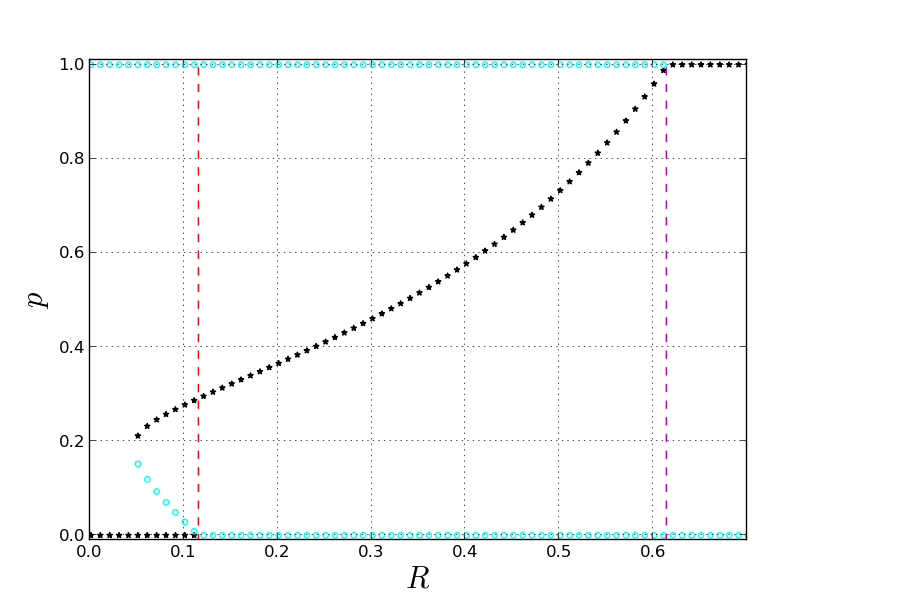}    
   \vspace{0.2cm}
   \includegraphics[width=0.5\textwidth]{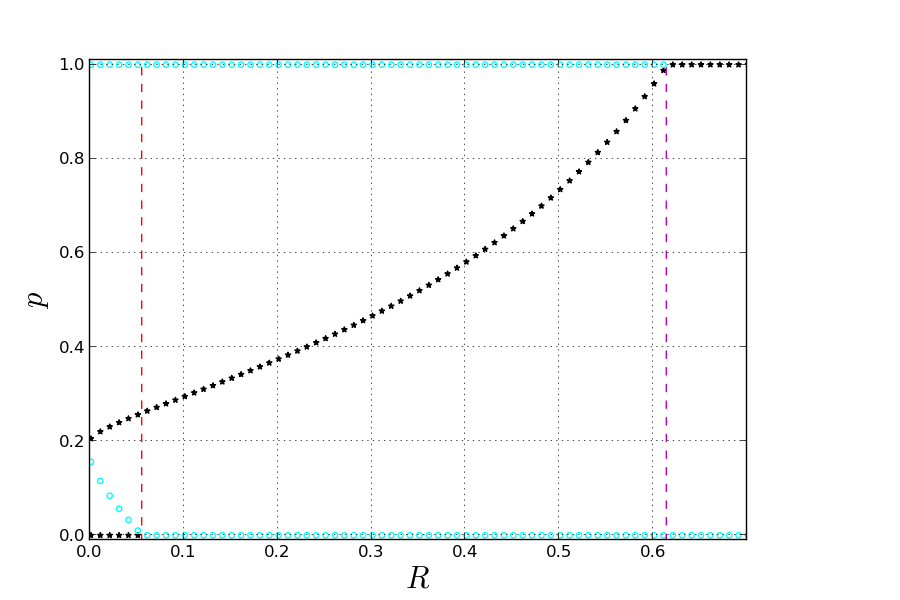} \hspace{-0.2cm} 
  \includegraphics[width=0.5\textwidth]{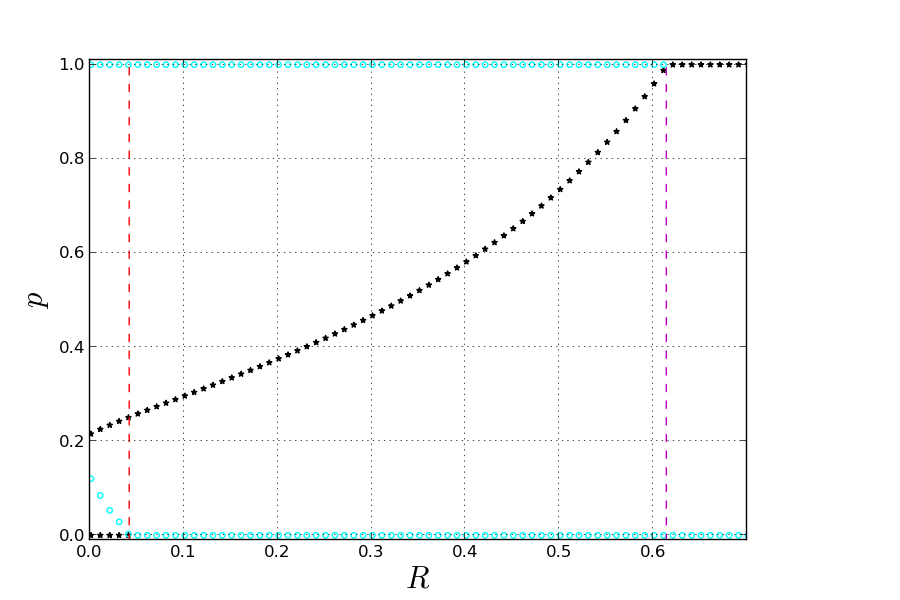}   
  \caption{{\bf THR}: $n=30$, $\theta = 10$, $C=1$, $A = T+1$, $A' = 2T$. 
  ({\bf Top Left}) $\bm {T=1} \ (\bm {A=2}, \, \bm{A^{\prime}=2})$:  
  $R_s=33.5\%$ ($m_s=3.2\%$) and  $R_f=61.4\%$ ($m_f=1\%$).
  ({\bf Top Right}) $\bm {T=10} \ (\bm {A=11}, \, \bm{A^{\prime}=20})$:
  $R_s=11.5\%$ ($m_s=10.7\%$) and  $R_f=61.4\%$ ($m_f=1\%$)
  ({\bf Bottom Left}) $\bm {T=100} \ (\bm {A=101}, \, \bm{A^{\prime}=200})$:
  $R_s=5.5\%$ ($m_s=20.2\%$) and  $R_f=61.4\%$ ($m_f=1\%$). 
  ({\bf Bottom Right}) $\bm {T=300} \ (\bm {A=301},  \, \bm{A^{\prime}=600})$:
  $R_s=4.2\%$ ($m_s=24.7\%$) and  $R_f=61.4\%$ ($m_f=1\%$).
  Open cian circles
  represent unstable fixed points, black stars represent stable fixed points. The critical relatedness for invasion
  by types A, $R_s$, is indicated by a red dashed line while the critical relatedness for invasion by types N, $R_s^N$, 
  is depicted as a magenta dashed line. The pictures show that here $R_f = R^N_s$.
  }
\label{fig:ITHRn=30}
\end{figure}

\newpage

\begin{figure}[ht!]
   \includegraphics[width=0.5\textwidth]{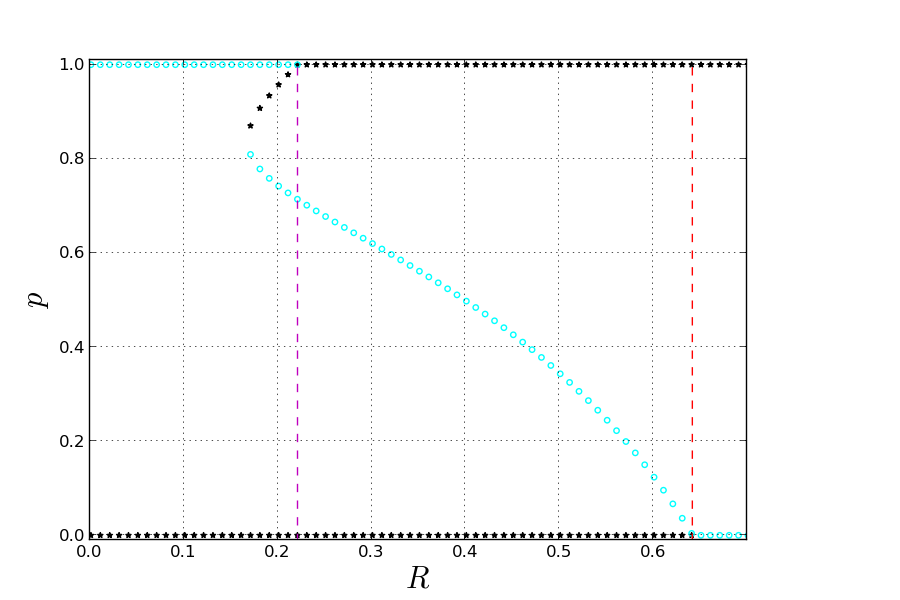} \hspace{-0.2cm} 
  \includegraphics[width=0.5\textwidth]{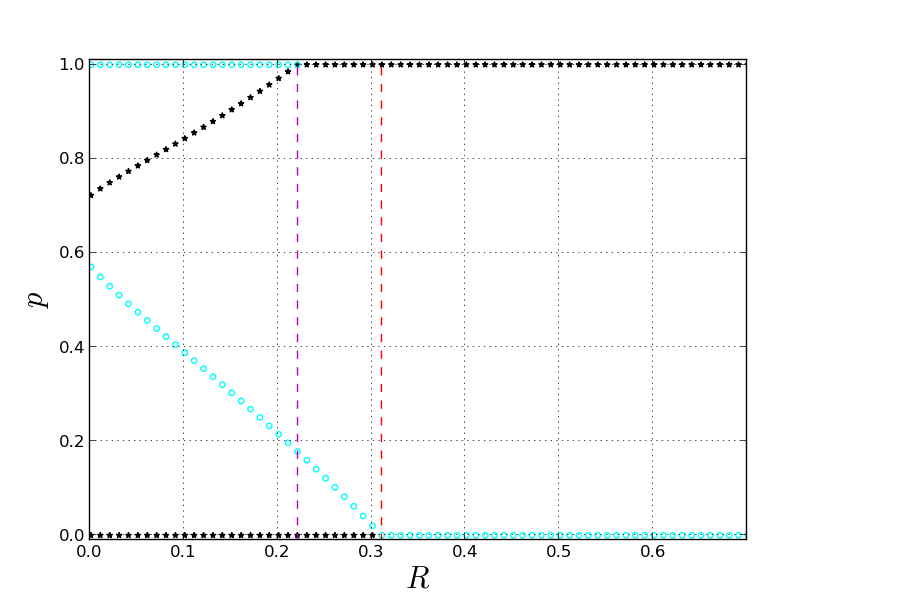}    
   \vspace{0.2cm}
   \includegraphics[width=0.5\textwidth]{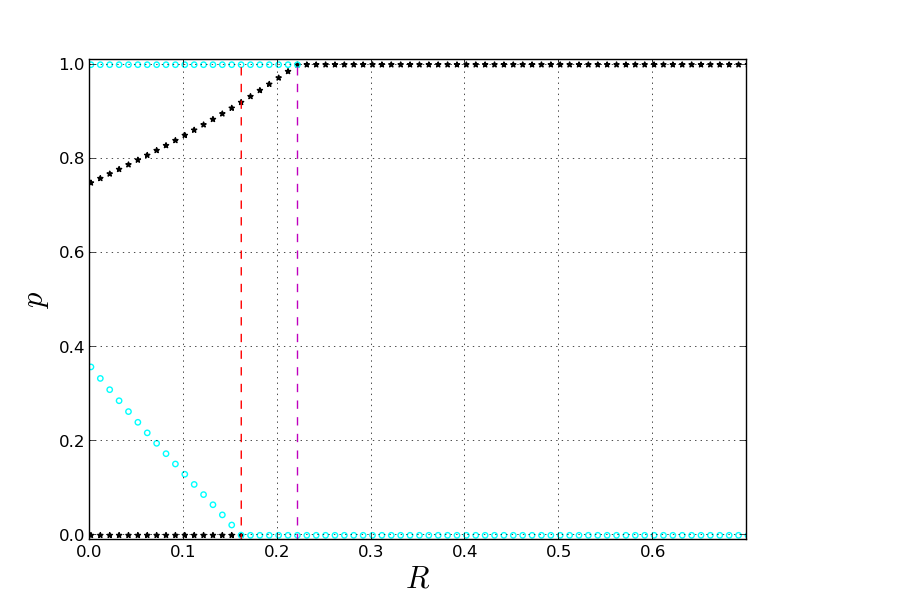} \hspace{-0.2cm} 
  \includegraphics[width=0.5\textwidth]{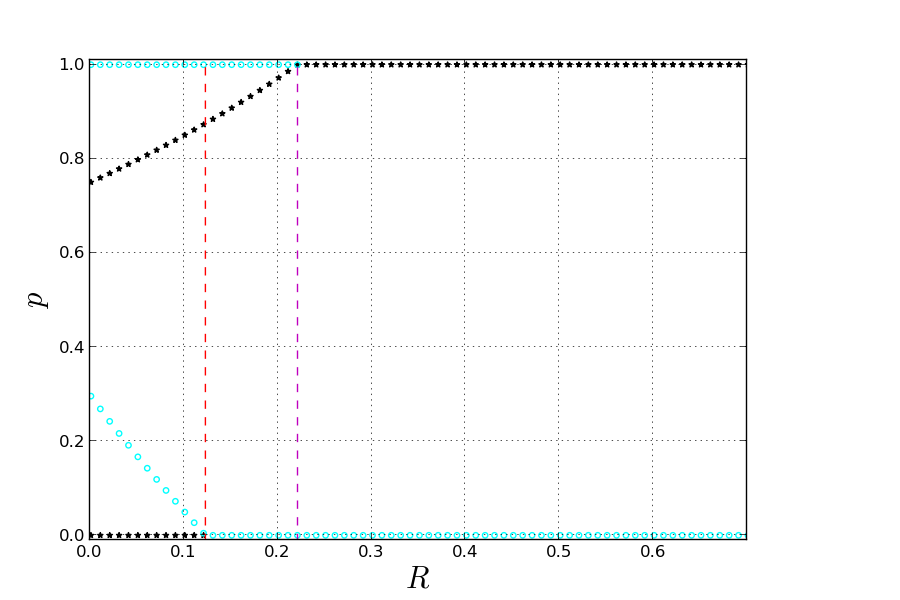}   
  \caption{{\bf THR}: $n=10$, $\theta = 8$, $C=1$, $A = T+1$, $A' = 2T$. 
  ({\bf Top Left}) $\bm {T=1} \ (\bm {A=2}, \, \bm{A^{\prime}=2})$:  
  $R_s=R_f=64.2\%$ ($m_s=m_f=2.7\%$) and $R^N_s=22.1\%$ ($m^N_s=14\%$).
  ({\bf Top Right}) $\bm {T=10} \ (\bm {A=11}, \, \bm{A^{\prime}=20})$:
  $R_s=R_f=31\%$ ($m_s=m_f=9.5\%$) and $R^N_s=22.1\%$ ($m^N_s=14\%$).
  ({\bf Bottom Left}) $\bm {T=100} \ (\bm {A=101}, \, \bm{A^{\prime}=200})$:
  $R_s=16.1\%$ ($m_s=18.9\%$) and $R_f=R^N_s=22.1\%$ ($m_f=m^N_s=14\%$). 
  ({\bf Bottom Right}) $\bm {T=300} \ (\bm {A=301},  \, \bm{A^{\prime}=600})$:
  $R_s=12.3\%$ ($m_s=23.6\%$) and $R_f=R^N_s=22.1\%$ ($m_f=m^N_s=14\%$).
  Open cian circles
  represent unstable fixed points, black stars represent stable fixed points. The critical relatedness for invasion
  by types A, $R_s$, is indicated by a red dashed line while the critical relatedness for invasion by types N, $R_s^N$, 
  is depicted as a magenta dashed line. The top pictures show $R_f = R_s$, while the bottom pictures show $R_f = R^N_s$.
  }
\label{fig:ITHRn=10,theta=8}
\end{figure}





\newpage

\section{Conclusions}
\label{sec:conclusions}
For a large class of group structured populations, involving competition within groups and
possibly also among groups, and/or allowing for elasticity in group size with local regulation, 
the equilibria under
weak selection are obtained from equating $\Delta p = 0$ in (\ref{Deltap}),
or equivalently, $D(p) = 0$ in (\ref{equilibria})
(with inputs from ((\ref{VA}), (\ref{VN}) and (\ref{stationarity})). 
The stability of each one of these equilibria is obtained from the sign of $\Delta p$, 
or equivalently, the sign of $D(p)$, close to it. 
In particular rare mutant alleles A will invade when (\ref{invasion}) holds.
If groups are large and the migration rate low, these conditions take simpler 
forms, provided by (\ref{VAlimit}), (\ref{VNlimit}) and (\ref{invasionlimit}),
or (\ref{invasionlimit+}), 
in which only the scaled gene flow parameter $mn_{\mbox{eff}}$ appears, 
or, equivalently, only the relatedness parameter $R = 1/(1+2mn_{\mbox{eff}})$ appears.  
All these conditions are easy to apply, providing tools that can address any 
sort of intra-group interaction and the complexities that result from multi-individual,
possibly iterated,
games accross a life-cycle, including contingencies of behavior
\cite{Joshi, BR, KJC-B, Aviles, HMND, GWW, Vee1, AS, 
GYO, CRL, SDZ,  
NTW1,
HP, BJRB,
BG, BGB}. 
In this way one can study biological models of social evolution in group-structured 
populations, in which gene action is non-additive, producing marginal fitness functions under
weak selection that are non-linear functions of group composition. 
Our methods can therefore
be used when approaches based on differentiability of fitness functions \cite{TF, Frank, WGF, GWW, Rou, LKWR, LR} 
are not applicable, as explained in \cite{SBV}.  
  
Under the two conditions: (C1) isolated mutants have lower fitness than the wild type; (C2) mutants that 
are in groups with no wild types have fitness that is larger than the wild types, 
the following regimes occur, when we start from a small fraction of mutants. 
(1) No invasion possible for high gene flow; (2) invasion leading to fixation for low 
gene flow; and possibly also: (3) invasion leading a polymorphic equilibrium for intermediary levels of gene flow. 
The three regimes are often present for iterated 
public goods games with contingent cooperation, and for threshold models, as we observed in Section \ref{sec:examples}. 
This contrasts to what happens with linear public goods games, for which there is no 
possibility of polymorphic equilibria under weak selection, suggesting that non-linearities
in fitness functions may be present when such equilibria are observed. (Compare with \cite{R-GGWG} 
where strong selection was proposed as an explanation.)  
Invasion of the cooperative types A can occur in these models under modest levels of group
relatedness, compatible with values observed in several species
\cite{HC} (Tables 6.4 and 6.5), 
\cite{MHam} (Table 4.9) and \cite{BG}. 
This result extends to a broad class of population structures the main conclusion in \cite{SVC},
showing that population viscosity, without kin recognition, can produce levels of genetic assortment 
that are sufficient for the spread of cooperative/altruistic intra-group behavior. In other words, 
contrary to a widespread claim
\cite{MS64, Williams, Ham75, MS76,   
Aoki, CA, Kimura,   
Leigh83, MS, 
Leigh, WMG} 
the biological conditions for ``the good of the group to override the interest of the individual'' are not stringent.
The opposite conclusion had been obtained and reinforced over the years based on the analysis of 
special models, mostly variants of a linear public goods game. The possibility of correcting that 
misunderstanding using the techniques from \cite{SVC} and the current paper highlights the importance of
having available good methods for analyzing non-linear marginal fitness functions.    

\

\noindent
{\bf Acknowledgments:}
The authors thank Clark Barrett, Nestor Caticha and Sarah Mathew 
for stimulating conversations and feedback on various 
aspects of this project. 
This project was partially supported by CNPq, under grant 480476/2009-8.



\section{Appendix A. Variable group size}
\label{sec:variablen}
The kind of population structure that is being considered in a simplified fashion in Gen.2
can be described as follows. There is a typical group size $n_0$, groups with fewer members 
tend to generate larger groups in the next generation (possibly because resources are abundant
for the small group), while groups with more members tend to generate smaller groups in the 
next generation (possibly because resources are scarce for the large group). This can be 
modeled by fitness functions 
$w^A_{n,k} = (1 + \delta v^A_{n,k}) h(n/n_0)$ and 
$w^N_{n,k} = (1 + \delta v^N_{n,k}) h(n/n_0)$, 
with a function $h(s)$ that is decreasing and takes the value 1 at $s=1$.    
If the typical group size, $n_0$, is large, the law of large numbers 
implies that each group that is created has size close to its expected value. In this case 
we should expect group sizes that are generally not far from $n_0$, in relative terms. 
A full analysis of this framework for moderate sizes of $n_0$ would nevertheless 
involve the evolution of group sizes. For this reason, we provided 
a more idealized, but also more tractable approach, in our generalization Gen.2 of 2lFW.    

We present now in detail an instance of the mathematical assumptions introduced 
in Gen.2, so that we can be sure that this can be done in a mathematically sound way. 
Our goal is primarily to be sure of the mathematical consistency of the assumptions
made in Gen.2, not aiming now for biological realism. 
The reader should also keep in 
mind that in the analyzis in the paper, details as those discussed next are not relevant.  
This robustness of the methods presented in the paper is one of their strengths. 
Our example will be mathematically as simple as possible. We suppose that  
each group creates in the average 1 group in the next generation and that groups 
always have one of the three sizes 
$n_0$, $n_0-1$, or $n_0+1$.
(The notation 
$n_0$ replaces the $n$ from the paper in this appendix, so that here we can use 
$n$ as a free variable to designate an arbitrary group size.)  
We will suppose that for the possible values of $n$ and $k$, the fitness functions are 
as presented in the previous paragraph.  
Therefore $\bar{w}_{n,k} = (1 + \delta \bar{v}_{n_0,k})h(n/n_0)$,
where $\bar{v}_{n,k} = (k v^A_{n,k} + (n-k) v^N_{n,k})/n$. 
The assumptions on the function $h$ imply that 
$\bar{w}_{n_0,k} = 1 + \delta \bar{v}_{n_0,k}$, 
and for small $\delta > 0$ we have 
$\bar{w}_{n_0-1,k} > 1$ 
and $\bar{w}_{n_0+1,k} < 1$, for all $k$. 

We want to check if the transition probabilities 
$\mbox{Pr} (n'|n,k)$
that give the distribution 
of the size $n'$ of an offspring group, conditioned on the size $n$ of its parental 
group and the number $k$ of members of that group that are types A, can satisfy the 
conditions proposed in Gen.2. For this we first write
$\mbox{Pr} (n_0|n_0,k) = 1 - \alpha_k -\beta_k$,
$\mbox{Pr} (n_0-1|n_0,k) = \alpha_k$ and
$\mbox{Pr} (n_0+1|n_0,k) = \beta_k$.
Then, for given fitness functions as above 
we must find probabilities $\alpha_k$ and $\beta_k$, both of order $\delta$, so that 
we obtain the proper average expected number of offspring of a group 
of size $n_0$ that includes $k$ types A. This means that we must solve 
$n_0(1-\alpha_k -\beta_k) + (n_0-1) \alpha_k + (n_0 + 1) \beta_k 
= 
n_0 (1 + \delta \bar{v}_{n_0,k})$.
This has many solutions when $\delta > 0$ is sufficiently small, 
the simplest one being $\alpha_k = - \delta n_0 \bar{v}_{n_0,k} $, 
$\beta_k = 0$ if $\bar{v}_{n_0,k} < 0$, 
and  
$\beta_k = \delta n_0 \bar{v}_{n_0,k} $, 
$\alpha_k = 0$ if $\bar{v}_{n_0,k} \geq 0$.
Biologically this means that a group of size $n_0$ typically creates only groups of size $n_0$,
but when the average fitness in the group is slightly larger (smaller) than 1, then the 
group creates with some small positive probability a group with one more (one less) member. 
Next we have to specify what $\mbox{Pr}(n'|n_0-1,k)$ and 
$\mbox{Pr}(n'|n_0+1,k)$ are. We only do it for the 
former, since the latter is analogous. 
We write $\mbox{Pr}(n_0-1|n_0-1,k) = 1-\gamma_k$, 
$\mbox{Pr}(n_0|n_0-1,k) = \gamma_k$. 
We have to be sure that, when $\delta$ is small,  
for each value of $k$, we can find probabilities $\gamma_k > 0$ so that  
$(n_0 - 1) \bar{w}_{n_0-1,k} = (1- \gamma_k) (n_0-1) + \gamma_k n_0$. 
This is equivalent to $\gamma_k = (n_0-1) (\bar{w}_{n_0-1,k} - 1)$. 
In case (and only in case) $\bar{w}_{n_0-1,k}$ is slightly larger than 1 for each value of $k$, the 
corresponding $\gamma_k$ are between 0 and 1, as 
we needed. This condition on $\bar{w}_{n_0-1,k}$ 
means that 
$1 < h((n_0-1)/n_0) < n_0/(n_0 - 1)$, 
and means that groups of size $n_0-1$ should be 
more productive than groups of size $n_0$, but imposes an upper bound on by how much.
An analogous condition holds on the other side:  
$n_0/(n_0+1) < h((n_0+1)) < 1$.  
(If these condition on $h$ fail, one needs to allow more values of $n$ to be reached in the model.)  

Having specified $\mbox{Pr}(n'|n,k)$ in the previous paragraph, and assuming 
that intragroup competition is modeled by Fisher-Wright sampling (i.e., we are not in the more general 
setting of Gen.4), we have the complete transition matrix:
\\ $\mbox{Pr}(n',k'|n,k) = \mbox{Pr}(n'|n,k) \, \mbox{bin}(k'|n', k w^A_{n,k}/ n \bar{w}_{n,k})$.
This provides us with a well defined stochastic process in which each individual in each 
generation has an expected number of offspring given by the proposed fitness functions
$w^A_{n,k}$ and $w^N_{n,k}$. (Keep in mind that migration among groups, after they are 
created, connects the $g$ groups, so that the stochastic process is a Markov chain, but only
in a very large state space involving the whole population. Under weak selection, though, we will 
be able to consider the migrants to a focal group as coming from a metapopulation with fixed 
frequency $p$ of types A, and in this way effectively reduce the analysis to that of a Markov chain 
acting on a single focal group.)  

There is a particularly nice twist to the framework under Gen.2. The fitness functions 
$w^A_{n,k}$ and $w^N_{n,k}$ in this setting can be absolute fitnesses 
(expected number of offspring of and individual), not just relative fitnesses. In 2lFW,
with fixed total population of size $ng$, the average absolute fitness must always be 1,
but this is not the case under Gen.2. And in the example in the previous two paragraphs,
$w^A_{n,k}$ and $w^N_{n,k}$ are indeed absolute fitnesses. 
(This is what we assumed, when we wrote equations that had to be 
safisfied by the transition probabilities.)
This fact can be puzzling at first 
sight. If $w^A_{n,k}$ and $w^N_{n,k}$ are absolute fitnesses and we have, say $v^A_{n,n} > 0$
and types A fixated (every individual is type A), then it would seem that everyone has 
an average fitness larger than 1 in a population in equilibrium, something that cannot 
happen. The ({\it incorrect}) reasoning behind this idea is that in equilibrium, with small 
$\delta> 0$, almost all groups are of size $n_0$ ({\it correct}) and that for this size 
$w^A_{n_0,n_0} = 1 + \delta v^A_{n_0,n_0} > 1$ ({\it correct}). So we seem to conclude that 
the average fitness in equilibrium must be larger than 1 by an amount of order $\delta$.  
But in this reasoning we are forgetting that even if only a fraction of order $\delta$ of 
groups in equilibrium have size $n_0 + 1$, individuals in these groups have fitness  
$w^A_{n_0 + 1, n_0+1} = (1+\delta v^A_n) h((n_0+1)/n_0) < 1$.
This provides a term to be added to the average fitness that is below 1 by an amount of 
order $\delta$; sufficient to explain how such an equilibrium is possible and has the 
necessary average absolute fitness 1. Most groups will have size $n_0$, 
and its members produce slightly more then one offspring.  
But the few groups that have size $n_0 + 1$ compensate this push upwards, since their members 
produce in the average a number of offspring below 1 by a fixed amount. 

It is also very intructive to compare the situation in which types N are fixated with that 
in which types A are fixated. Under the assumption 
that $v^N_{n,n} = 0$ and, as above, $v^A_{n,n} > 0$. In each one of these fixated equilibria, 
the average absolute fitness is by necessity 1. But the average size of the population is 
larger in the case A is fixated. When N is fixated groups only have size $n_0$ in equilibrium,
while when A is fixated groups have size $n_0$ or $n_0 + 1$ in equilibrium. In this case
only a fraction of order $\delta$ of groups have size $n_0 + 1$, so that the average group 
size exceeds 1 only by an amout of that order, but it does exceed it.      
  

\section{Appendix B. General intragroup transition matrix}
\label{sec:intragroup}
Here we provide some details on Gen.4. First note that if one is in the setting of Gen.2, or Gen.3,  
then one would in principle have to define transition probabilities from $(n,k)$ to 
$(n',i)$, and that the distribution of $i$ may depend on $n$ in addition to $k$. But 
because we are only concerned with weak selection in this paper, we will only need to 
consider the transition matrix in case $\delta=0$ (small $\delta$ is treated as a 
perturbation), and in this case, even under Gen.2 and Gen.3, we are restricted to groups of a 
fixed size $n$. For this reason, we use the notation $P_{k,i}$ 
for the $\delta = 0$ case. Note that $P_{k,i}$ is a Markov transition matrix, 
and recall that we are assuming $\sum_i iP_{k,i} = k$ 
(when $\delta = 0$, A and N are neutral markers)
and $P_{0,0} = P_{n,n} = 1$
(the states $0$ and $n$ are traps for the Markov chain generated by $P_{k,i}$).
 
Under the generalization Gen.4, (added possibly to Gen.1, Gen.2 and Gen.3) 
the Markov transition matrix 
$T_{j,k}$, that appears 
in (\ref{recursionforf}) 
and (\ref{stationarity}) 
takes the form

\be
T_{j,k} \ = \ \sum_{i, i', i''} \, 
P_{j,i} \, 
\mbox{bin} (i'|i,m) \, 
\mbox{bin} (i''|n-i,m) \, 
\mbox{bin} (k- i + i'| i' + i'', p).
\label{Tjk}
\ee
Explanation: with probability $P_{j,i}$, 
the new group is created with $i$ types A and $n-i$ types N, 
of which, respectively, $i'$ and $i''$ migrate and are replaced with 
migrants from the metapopulation. The number $k$ of types A in the group after migration is then 
the sum of $i-i'$ (non-migrants) and a binomial random variable corresponding to 
$i' + i''$ attempts, each with probability $p$ of success (migrants). 
(We use the standard convention that $\mbox{bin}(i|n,q)= 0$, when 
$i < 0$ or $i > n$.)

All the claims in Sections \ref{sec:drift} and \ref{sec:ws} hold, with the same arguments 
given there and with $\varphi(p)$ being the (unique once $m > 0$) 
stationary distribution of the Markov chain with transition matrix $T_{j,k}$ given in (\ref{Tjk}).
Also the statements in Appendices C, D and E hold without changes in the argumentation.

We present now two types of examples of intragroup transition 
with substantially different biological meanings. 
In one important class of examples, each individual in the parental group can be 
thought of as creating an independent random number of offspring with a certain distribution 
(independent of $n$) with mean proportional to individual fitness, 
conditioned on the total number of offspring created being $n$ (or whatever the group size is 
in case of Gen.2). When the individual offspring distribution
is Poisson we have the Fisher-Wright case. We will refer to these examples as the case of ``distributed'' 
production (or transition) scheme. 

In contrast, one member of the parental group can be chosen at random, with probability 
proportional to individual fitness, and mother the whole new group.
This example, with high reproductive skew, will be referred to as ``concentrated''
production (or transition) scheme.

The distributed and concentrated intragroup transition schemes 
illustrate different aspects of the generalized 2lFW 
framework. In the concentrated case, we can easily express $P_{k,i}$ and use it to 
compute the corresponding $T_{j,k}$ and $\varphi_k(p)$. This is not the case in the 
distributed case, for which $P_{k,i}$ is usually difficult to compute. But in 
contrast to the concentrated case, the distributed case 
(supposing that the offspring distribution for the individuals has a finite second moment)
satisfies the conditions 
for Wright's beta approximation (\ref{limbeta}) to $\varphi_k(p)$ to apply. This is so 
because in this case, when $n$ is large, there is sufficient independence among 
the types of the members of the offspring group (conditioned on the fraction of 
types A in the parental group), for the diffusion approximation used to derive that 
beta distribution to be applicable. This adds significantly to the relevance 
of the Fisher-Wright case and the beta approximation results in Section \ref{sec:nm}.               
They represent in good approximation a broad class of intragroup transition 
schemes, when $n$ is large.

\section{Appendix C. Graphical construction of the equilibria}
\label{sec:partitions} %
The equilibria $\varphi(p) = (\varphi_0(p), ..., \varphi_n(p))$ can be easily computed using 
(\ref{stationarity}). 
(This is not so easy in case of Gen.4, but can still be done using  
(\ref{Tjk}).) 
This equation can also be used to derive the properties of $\varphi(p)$. Here we present an alternative
way of describing these equilibria, that will appeal to readers who are fond of graphical devices, 
and that provides additional intuition. 

The idea is to partition the members of a group into classes of equivalence defined by the property 
of being IBD. In each one of these (IBD)-classes, all the individuals will be of type A, or all will be 
of type N, with respective probabilities $p$ and $1-p$, independently from class to class. 

To obtain the (IBD)-classes, we should follow 
the lineages of the members back in time
and record coalescence events and migration events. Once 
there is a migration event in a lineage, we can stop following this lineage.
The migration even means that this lineage reached the group in that generation. All 
members of the (IBD)-class 
of descendents from this migrant will therefore be of the same type as
this migrant. Because we are considering equilibrium, with frequency $p$ of types A, 
this type will be A or N with probability $p$ or $1-p$. And because the population is
very large and migration is random, the different migrants to a group are type A or N 
independently of each other, implying that the IBD-classes are type A or N independently 
of each other.    

The benefit of looking at the equilibria $\varphi(p)$ in the fashion above, is that it allows one
to see what happens in special cases, especially extreme cases, easily. 
When $p$ is close to $0$ (or $1$), all the (IBD)-classes will be likely to have individuals 
who are type N (or type A). This explains (\ref{varphiaspto0}) and  (\ref{varphiaspto1}).
When $m=1$ all individuals 
are migrants in generation $t$. Therefore each individual is in a different (IBD)-class, and clearly 
(\ref{varphim1}) holds. In the opposite extreme, when $m \to 0$, all the lineages of the members of
the group are likely to coalesce before migration, we obtain then, with overwhelming probability, 
a single (IBD)-class. With probability $p$, this class will contain only types A, while with probability $1-p$ 
it will contain only types N. This explains (\ref{varphiasmto0}). 

\section{Appendix D. Relatedness}
\label{sec:rel} %
Select a group at random in generation $t$, and from this group select in order two distinct
individuals (the focal and the co-focal). Define $\hat p$ as the 
fraction of types A in that group, 
$\kappa$ as the fraction of types A among the $n-1$ individuals in this group that are distinct
from the focal individual (this is the focal's social environment), 
$A_1$ as the event that the focal individual is type A, 
$A_2$ as the event that the co-focal individual is type A. 
Define $I_i$, $i=1,2$ as the random variable that takes value 1 when $A_i$ happens and value 0 otherwise.
Say also that the focal and the 
co-focal individuals are identical by descent (IBD), if following their lineages back in time,
they coalesce before either one experiences a migration event.

Relatedness can be defined as certain regression coefficients, namely the regression of $I_2$ on $I_1$ 
or, equivalently, that of $\kappa$ on $I_1$:
\begin{eqnarray}
R^{\mbox{reg}} \ & = & \ \beta_{I_2,I_1} \ = \ 
\frac{\mbox{Cov}(I_1, I_2)}{\mbox{Var}(I_1)} \ = \ \E(I_2|A_1) - \E(I_2|A_1^c)  
\nonumber \\ \ & = & \
\E(\kappa|A_1) - \E(\kappa|A_1^c) \ = \  \frac{\mbox{Cov}(I_1, \kappa)}{\mbox{Var}(I_1)} \ = \
\beta_{\kappa, I_1}. 
\label{Rregression}
\end{eqnarray}
Here the first, the second and the last equalities are definitions, and the others are elementary 
probability identities. 
Alternatively, relatedness can be defined by
\be
R^{\mbox{IBD}} \ = \ \mbox{Pr} \, (\mbox{focal and co-focal are IBD}). 
\label{RIBD}
\ee
Wright's $F_{ST}$ statistics can be defined by 
\be
F_{ST} \ = \ \frac{\mbox{Var}(\hat{p})}
{\mbox{Var}(I_1)}.
\label{FST}
\ee
When the sampling of the focal and co-focal is done in one of the equilibria $\varphi(p)$, 
the following well known relationships hold among the
three definitions above and (\ref{R}):
\be
R \ = \ 
R^{\mbox{IBD}} \ = \ R^{\mbox{reg}} \ = \ \frac{n F_{ST} - 1}{n-1}.
\label{Rs}
\ee
For the reader's benefit, we present short derivations next. 

The first equality in
(\ref{Rs}), is a consequence of the following equilibrium recursion. The focal 
and co-focal will be IBD if and only if neither one is a migrant 
(probability $(1-m)^2$)
and they either
have the same mother (probability $1/n_{\mbox{eff}}$), or they have distinct mothers that are 
IBD (probability $(1 - (1/n_{\mbox{eff}})) R^{\mbox{IBD}}$). Hence
$R^{\mbox{IBD}} = (1-m)^2 ((1/n_{\mbox{eff}}) + (1-(1/n_{\mbox{eff}})) R^{\mbox{IBD}}$,
from which we get $R^{\mbox{IBD}} = R$, given by (\ref{R}).

To derive the second equality in (\ref{Rs}), let $D$ be the event that focal and co-focal 
are IBD. Because we are sampling from the equilibrium $\varphi(p)$, we know that the population
has been in this equilibrium for several ($>> 1/m$) generations prior to the sampling. Therefore,
$\mbox{Pr}(A_1 A_2 | D) = p$ and $\mbox{Pr}(A_1 A_2 | D^c) = p^2$. (We use the standard convention 
of writing $A \cap B$ simply as $AB$. One way to thing about these conditional probabilities
is in terms of the (IBD)-classes defined in Appendix C. %
The even $D$ means that the focal and 
cofocal individuals are in the same (IBD)-class. In this case they are of the same type, while in 
the opposite case, they are of independent types.)
Hence $\mbox{Cov}(I_2,I_1) = \mbox{Pr}(A_1 A_2) - p^2 = p R^{\mbox{IBD}} + p^2 (1 - R^{\mbox{IBD}}) -p^2
= p(1-p)R^{\mbox{IBD}} = \mbox{Var}(I_1) R^{\mbox{IBD}}$, which yields the second equality in (\ref{Rs}).
 
The third equality in (\ref{Rs}), does not depend on having sampled in equilibrium, and 
results from the following elementary probability identity:
\be
\mbox{Var}(\hat{p}) \ = \ \frac{n \mbox{Var}(I_1) \, + \, n(n-1) \, \mbox{Cov}(I_1,I_2)}{n^2}
\ = \ \frac{\mbox{Var}(I_1) \, (1 + (n-1) R^{\mbox{reg}})}{n},
\label{forFST}
\ee
where we used (\ref{Rregression}) in the second step.
Using the definition (\ref{FST}), this becomes $F_{ST} = (1 + (n-1) R^{\mbox{reg}}) / n$, which 
is equivalent to the last equality in (\ref{Rs}).

\section{Appendix E. Condition for invasion based on IBD distribution}
\label{sec:pi} 
Display (2) in \cite{SVC} implies that under weak selection, the condition for invasion 
by allele A (instability of the equilibrium with $p=0$) can be written as 
\be
\sum_{k=1}^n \pi_k v^A_k \ > \ 0, 
\label{invasionpi}
\ee   
where $\pi$ is the stationary distribution of the Markov chain on 
$\{ 1, ..., n \}$, with transition matrix  
\be
Q_{i,j} =  
m \, 1_j 
\ + \ (1-m) \, \mbox{bin}(j-1 \, | \, n-1 \, , \, (1-m)i/n),
\label{Q}
\ee
with notation: $1_j = 1$ if $j = 1$ and $1_j = 0$ if $j \not= 1$. Therefore $\pi$ can be computed 
from the stationarity condition $\pi = \pi Q$ and the normalization condition $\sum_{k=1}^n \pi_k = 1$. 

The distribution $\pi$ has a very simple and biologically natural interpretation,
that we recall next. 
Two individuals are said to be IBD if following their lineages 
back in time, they coalesce before a migration event affects either one.
Then, as we explain in \cite{SVC}, $\pi_k$ is the probability that if we choose 
at random a focal individual, it will have exactly $k$ individuals in its group 
that are IBD to it (self included). For this reason, we refer to $\pi$ as the IBD 
distribution. 

Comparing (\ref{invasion}) with (\ref{invasionpi}), we see that since $v^A_k$ is 
arbitrary, the term inside parentesis in (\ref{invasion}) must equal $\pi_k$. 
But the computation of $\pi$ as the stationary distribution of $Q$ is more 
straightforward. For this reason we used (\ref{invasionpi}) to compute the 
values of $m_s$ and $R_s$ in our examples in Section \ref{sec:examples}.
This is done by replacing (\ref{invasionpi}) with the corresponding equality
and finding the value of $m = m_s$ that solves this equation. (Recall 
that $R_s$ is then defined by pluging in $m = m_s$ in (\ref{R})). 

In the same way that one asks for conditions for the stability of the equilibrium
with $p=0$, one can ask the analogous question about the equilibrium with $p=1$. 
When can allele N invade a population in which there are only types A? 
We define $m_s^N$ (and the corresponding $R_s^N$) as the least amount of migration 
(largest degree of relatedness) for which this invasion can happen. Since in the 
derivations of (\ref{invasion}) and (\ref{invasionpi}), there is nothing that 
qualitatively distinguishes types A from types N, we can use these invasion 
conditions, with the appropriate quantitative modifications, to compute $m_s^N$ and $R_s^N$.
The version of the condition for invasion of allele N based on (\ref{invasionpi}) reads
\be
\sum_{k=1}^n \pi_{n-k} v^N_k \ > \ v^A_n. 
\label{invasionNpi}
\ee    
The only subtlety is the fact that the right-hand-side in (\ref{invasionpi}) is $v^N_0 = 0$,
that has to be replaced with $v^A_n$ in (\ref{invasionNpi}). Indeed, the meaning of the left-hand-side
in (\ref{invasionpi}) is the average marginal fitness of a focal type A (since it is rare, only 
individuals who are IBD to it in its group are also types A). And the meaning of the right-hand-side
in (\ref{invasionpi}) is the average marginal fitness of a focal type N, in the population dominated by 
types N and with few invading types A. That fitness is essentially $v^N_0$, since almost all types N 
then are in groups with no types A. This makes (\ref{invasionpi}) intuitive and an analogous reasoning 
with the roles of A and N interchanged makes (\ref{invasionNpi}) equally intuitive. 
The value of $m^N_s$ is now obtained from solving for equality in (\ref{invasionNpi}), and  
$R^N_s$ is then defined by pluging in $m = m^N_s$ in (\ref{R})).
 
\section{Appendix F. Linear public goods game and relatedness}
\label{sec:PG} %
In the basic example of the PG,
$v^A_n = -C + B(k-1)/(n-1)$ and
$v^N_k = Bk/(n-1)$, it is well known \cite{Ham75} that (\ref{Deltap}) takes the 
simple form 
\be
\Delta p \ = \ \delta \, p(1-p) \, (-C + BR) \ + \ O(\delta^2).
\label{Hamilton}
\ee 
Hamilton \cite{Ham75} derived (\ref{Hamilton}) using the Price equation and the relationship
between $R$ and Wright's $F_{ST}$ statistics. For completeness, and for the reader's benefit,
we present next two alternative derivations of (\ref{Hamilton}). 

A simple way to derive (\ref{Hamilton}) from (\ref{Deltap}) is to directly
compute $V^A(p) - V^N(p)$, using their definitions, (\ref{VA}) and (\ref{VN}).
Using also (\ref{mean}), we obtain, after some standard algebraic manipulations,
\begin{eqnarray}
V^A(p) - V^N(p) \ & = & \ 
-C \ + \ B \, 
\frac{
n \left [
\frac{
\sum_k (k/n)^2 \varphi_k(p) \, - \, p^2
}
{p(1-p)} 
\right ] 
\, - \, 1}{n-1}   
\nonumber
\\
\ & = & \ 
-C \ + \ B \, 
\frac{
n F_{ST}
\, - \, 1}{n-1}
\ = \ -C \ + \ BR,
\label{HamiltonFST}
\end{eqnarray}
where we used (\ref{Rs}) in the last step.

An alternative derivation of (\ref{Hamilton}) from 
(\ref{Deltap})
is to 
observe that the payoff to the focal individual can be written as $-C I_1 + B \kappa$, and  
therefore 
\begin{eqnarray}
V^A(p) - V^N(p) \ & = & \ \E(-C + B \kappa | A_1) \, - \, \E(B \kappa | A_2) 
\nonumber \\ \ & = & \ 
-C \, + \, B \, [\E(\kappa|A_1) - \E(\kappa| A_2)] 
\nonumber \\ \ & = & \ 
-C \, + B \, R^{\mbox{reg}} \ = \ -C + BR, 
\label{Hamiltonreg}
\end{eqnarray}
where we used (\ref{Rregression}) and (\ref{Rs}).



\end{document}